\documentclass[12pt,preprint]{aastex}
\def\etal{{\it et al.\ }}
\def\etall{{\it et al.}}
\def\deg{^\circ}  

\begin{document}

\begin{center}
\bf \LARGE
Three-Body Capture of Irregular Satellites: Application to Jupiter\\[50pt]
\end{center}

\medskip

\begin{center}
{\large Catherine Philpott${^1}$,  Douglas P. Hamilton${^1}$, and Craig B. Agnor$^2$}\\[20pt]

$^1$ Department of Astronomy, University of Maryland,
College Park, MD 20742-2421\\
E-mail: cphilpott@astro.umd.edu\\
hamilton@astro.umd.edu\\[10pt]

$^2$ Astronomy Unit, School of Mathematical Sciences, 
Queen Mary University of London, London, UK E14NS\\
E-mail: c.b.agnor@qmul.ac.uk\\[100pt]

Submitted to {\it Icarus}: October 19, 2009\\

\end{center}

\newpage

\begin{abstract}

  We investigate a new theory of the origin of the irregular
  satellites of the giant planets: capture of one member of a
  $\sim$100-km binary asteroid after tidal disruption.  The energy
  loss from disruption is sufficient for capture, but it cannot
  deliver the bodies directly to the observed orbits of the irregular
  satellites.  Instead, the long-lived capture orbits subsequently
  evolve inward due to interactions with a tenuous circumplanetary gas
  disk.

  We focus on the capture by Jupiter, which, due to its large mass,
  provides the most stringent test of our model.  We investigate the
  possible fates of disrupted bodies, the differences between prograde
  and retrograde captures, and the effects of Callisto on captured
  objects.  We make an impulse approximation and discuss how it allows
  us to generalize capture results from equal-mass binaries to
  binaries with arbitrary mass ratios.

  We find that at Jupiter, binaries offer an increase of a factor of
  $\sim$10 in the capture rate of 100-km objects as compared to single
  bodies, for objects separated by tens of radii that approach the
  planet on relatively low-energy trajectories.  These bodies are at
  risk of collision with Callisto, but may be preserved by gas drag if
  their pericenters are raised quickly enough.  We conclude that our
  mechanism is as capable of producing large irregular satellites as
  previous suggestions, and it avoids several problems faced by
  alternative models.

\end{abstract}

\section{Introduction}

\label{intro}

\subsection{Previously suggested capture models}

\label{models}

With discoveries accelerating in the last decade, we now know of over
150 satellites orbiting the giant planets.  About one-third of these
are classified as regular, with nearly circular and planar orbits.  It
is thought that these satellites are formed by accretion in
circumplanetary disks.  The majority of the satellites, however, are
irregular and follow distant, highly eccentric and inclined paths.  It
is widely believed that irregular satellites originated in
heliocentric orbits and were later captured by their planets, but the
details of how this occurred are still uncertain.  At least seven
different models have been proposed, involving dissipative forces,
collisions, resonances, and three-body effects.  Each model has its
own strengths and weaknesses.

In one long-standing theory, planetesimals are slowed as they punch
through the gas disk surrounding a young, growing planet (Pollack
\etall, 1979).  For this mechanism to be efficient, the gas must be
sufficiently dense to capture the planetesimals in one pass.  This is
problematic, however, because if the gas disk does not rarefy
substantially in $\sim$100-1000 years, the orbits of the new
satellites will decay inward, leading to collisions with the planet
and its regular satellites.  Furthermore, the atmospheres of Uranus
and Neptune have only a few Earth-masses of hydrogen and helium at
present, so their gas disks could not have been as extensive or
long-lived as those of Jupiter and Saturn.  A likely outcome of this
model, then, is that satellite capture should have been different at
Jupiter and Saturn than at Uranus and Neptune; however, current
observational estimates suggest equal efficiencies (Jewitt \&
Sheppard, 2005).  With a model similar to that of Pollack \etall,
{\'C}uk \& Burns (2004a) found that Jupiter's largest irregular
satellite, Himalia, would evolve inward to its current orbit in
$10^4-10^6$ years.  This tenuous gas, however, may make capture
difficult.

In another model, planetesimals are captured when the mass of the
planet increases (Heppenheimer \& Porco, 1977).  This mass growth
causes the planet's escape velocity to increase, rendering a
previously free planetesimal bound to the planet.  For this method to
be effective, the planet's mass must increase substantially on
$\sim$100-1000-year timescales.  However, in most planet formation
models (e.g. Pollack \etall, 1996), giant planet growth is
hypothesized to take place on timescales many orders of magnitude
longer than required by this capture scenario.  Furthermore, Uranus
and Neptune's gas deficiency implies that their growth was of very
short duration.  Thus, our current understanding of planetary
formation makes this model improbable.

The observation that the four giant planets contain approximately the
same number of irregular satellites (accounting for observational
biases; Jewitt \& Sheppard, 2005) has led to a renewal of interest in
capture theories that do not depend strongly on the planet's formation
process.  In one such scenario, a planetesimal collides with a current
satellite or another planetesimal in the vicinity of the planet,
resulting in its capture (Colombo \& Franklin, 1971).  Though
collisions were certainly more common in the early Solar System than
they are today, if they resulted in enough energy loss to permit
capture, they would likely also have catastrophically disrupted the
bodies.  Nevertheless, the fragments might then have become
independent satellites.

A fourth suggestion involves the possible instability in the orbits of
the outer planets early in the Solar System (e.g., by a 2:1 resonance
crossing between Jupiter and Saturn).  Outlining the theory of the
Nice model of Solar System evolution, Tsiganis \etal (2005) have shown
that such an event could cause Uranus and Neptune to have many close
approaches with each other and with Jupiter and Saturn.  During these
encounters, the influence of the massive interloping planet can cause
planetesimals to be stabilized as satellites (Nesvorn{\'y} \etall,
2007).  This method is promising but has an important disadvantage in
that Jupiter (and Saturn, to a lesser extent) sustains very few close
encounters relative to the ice giants.  Thus the gas giants are
inefficient at capturing satellites in this way (Nesvorn{\'y} \etall,
2007).

Astakhov \etal (2003) examined low-energy orbits that linger near
Jupiter and Saturn.  While these bodies are not permanently captured,
the authors found that some of them were stable for thousands of
years, long enough to allow a weak dissipative force such as gas drag
to complete the capture process.  However, the overall percentage of
temporary captures that do not escape is small, and many of these
bodies are threatened by collision with the planets' large outer
satellites (e.g., Callisto and Titan).

Agnor \& Hamilton (2006a) examined the capture of Triton from an
exchange reaction between a binary pair and Neptune.  Their motivation
stemmed from the newly-discovered abundance of binaries in small-body
populations.  Currently, it is estimated that binaries account for
$\sim$30\% of Kuiper belt objects (KBOs) with inclinations $<$
5$\deg$, $\sim$5\% of the rest of the KBOs (Noll \etall, 2008), and
$\sim$2\% of large main belt asteroids (diameters $>$ 20 km;
percentage increases for smaller objects; Merline \etall, 2007).  In
Agnor \& Hamilton's capture model, a binary is tidally disrupted and
one of its members, Triton, is captured as a satellite.  This process
is most effective for large satellites like Triton, with radius 1350
km.  However, the largest of the other irregular satellites are more
than 10 times smaller than Triton: Himalia at Jupiter is $\sim$85 km
in radius, Saturn's largest irregular, Phoebe, is $\sim$110 km,
Uranus's Sycorax is $\sim$80 km, and Neptune's Halimede and Neso are
only $\sim$30 km each.  Capturing these satellites via binary exchange
reactions would be significantly more difficult, as we will discuss
further below.

Finally, Vokrouhlick{\'y} \etal (2008) examined binary exchange
reactions during the first 100 Myr after an assumed Jupiter/Saturn 2:1
resonance crossing, using results of the Nice model (Tsiganis \etall,
2005) to guide their initial conditions.  Because planetesimal speeds
relative to the planets are high after the scattering phase of the
Nice model, they found that captures from binaries during that time do
not match current orbital parameters and occur too infrequently to
account for today's populations.

\subsection{Our model: Capture from 100-km binaries}

\label{ourmodel}

All of the above models have promising aspects coupled with important
limitations.  In this work, we seek to combine the best features of
several models into a viable capture scenario.  In particular, we
examine binaries (as in Agnor \& Hamilton, 2006a and Vokrouhlick{\'y}
\etall, 2008) as a way to augment capture from low-velocity orbits
resulting from three-body interactions like those studied by Astakhov
\etal (2003).  While Vokrouhlick{\'y} \etal (2008) studied exchange
reactions in the context of an assumed initial planetesimal
population, we focus on assessing the viability of the mechanism
itself.  Our goal is to determine how various parameters of the model
affect its plausibility.  We examine its viability at Jupiter, as a
number of the above models suggest that capturing at the largest gas
giant is especially difficult.

As the largest of the existing irregular satellites are $\sim$80-110
km, capture of objects in this size-range is particularly interesting.
Since it is likely that the irregular satellite population contains
collisional families (Nesvorn{\'y} \etall, 2003; Sheppard \& Jewitt,
2003), it may be the case that only the largest objects were captured,
while the smaller satellites formed later, via collisions.  For this
reason, we focus our investigation on capturing the $\sim$100-km
progenitors.

In order to stabilize and shrink the resulting capture orbits, a
dissipation source is required; we suggest a tenuous version of the
gas drag originally proposed by Pollack \etal (1979).  Two of
Jupiter's irregular satellites, Pasiphae and Sinope, as well as
Saturn's satellite, Siarnaq, and Uranus' Stephano, are found in
resonances that seem to require just such a weak dissipative force
(Whipple \& Shelus, 1993; Saha \& Tremaine, 1993; {\'C}uk \& Burns,
2004b).

Furthermore, a tenuous circumplanetary disk is consistent with current
theories of late-stage planetary formation.  Jupiter's massive gaseous
envelope of hydrogen and helium necessitates that it formed in the
Solar System's circumstellar gas disk.  Before the end of its
accretion, Jupiter was likely able to open a gap in the local density
distribution of the gas (for a review, see e.g. Papaloizou \etall,
2008).  After gap opening, gas continues to leak into the planet's
Hill sphere through the $L_1$ and $L_2$ points, but at a rate much
reduced in comparison to the previous epochs.  A tenuous
circumplanetary gas disk results (e.g., Lubow \etall, 1999; D'Angelo
\etall, 2003; Bate \etall, 2003), from which material may condense and
regular satellites may accrete near the planet (e.g., Canup \& Ward,
2002; Mosquiera \& Estrada, 2003).

In {\'C}uk \& Burns' study (2004a) of the Himalia progenitor's orbital
evolution, they considered circumjovian nebular conditions consistent
with hydrodynamical simulations of Jupiter's gap opening in a
circumstellar gas disk (e.g., Lubow \etall, 1999) and found that the
post-capture timescale for evolving this progenitor to its present
orbit to be roughly in the range of $10^4-10^6$ years.  This is
similar to the timescale in which extrasolar circumstellar disks
transition from optically thick to thin ($\sim10^5$ years; Skrutskie
\etall, 1990; Silverstone \etall, 2006; Cieza \etall, 2007).  The
similarity of timescales suggests that satellites captured at the
onset of disk dispersal have a good chance of experiencing stabilizing
orbital evolution while also avoiding collision with the planet.

The timescale for binary capture is very short compared to evolution
timescales from a tenuous gas disk.  Therefore, we focus our study
first on characterizing the effectiveness of binary capture in the
absence of gas.  In the following sections, we critically evaluate our
model for capturing irregular satellites from low-mass ($\sim$100-km)
binaries.  We begin with a closer examination of the three-body
capture process and then explore parameter space with a large suite of
numerical simulations.  We then discuss the ability of gas drag to
stabilize post-capture orbits in Section~\ref{survivability}.

\section{Three-body capture process}

\label{3bodycapture}

Binary capture first requires a close approach between a binary pair
and a planet.  As the pair approaches the planet on a hyperbolic
trajectory, its two components also orbit their mutual center of mass
(CM).  Hence, each member's speed with respect to the planet is a
vector sum of its CM speed ($v_{CM}$) and its orbital speed around the
CM.  If the binary passes close enough to the planet, it will be
tidally disrupted.  Following Agnor \& Hamilton (2006a), we make an
'impulse approximation' and assume that disruption is instantaneous,
so that the distance at which tidal disruption occurs ($r_{td}$) can
be estimated as:

\begin{equation}
\label{rtd}
r_{td} \approx a_{B}\left(\frac{3M_{P}}{m_{1} + m_{2}}\right)^{1/3} ,
\end{equation}

\noindent where $a_{B}$ is the semi-major axis of the binary, $M_{P}$
is the mass of the planet, and $m_{1}$ and $m_{2}$ are the masses of
the binary pair.  This tidal disruption radius is the distance to the
planet at which the binary's Hill sphere is no longer larger than the
binary itself.

\begin{figure}[htp]
\plotone{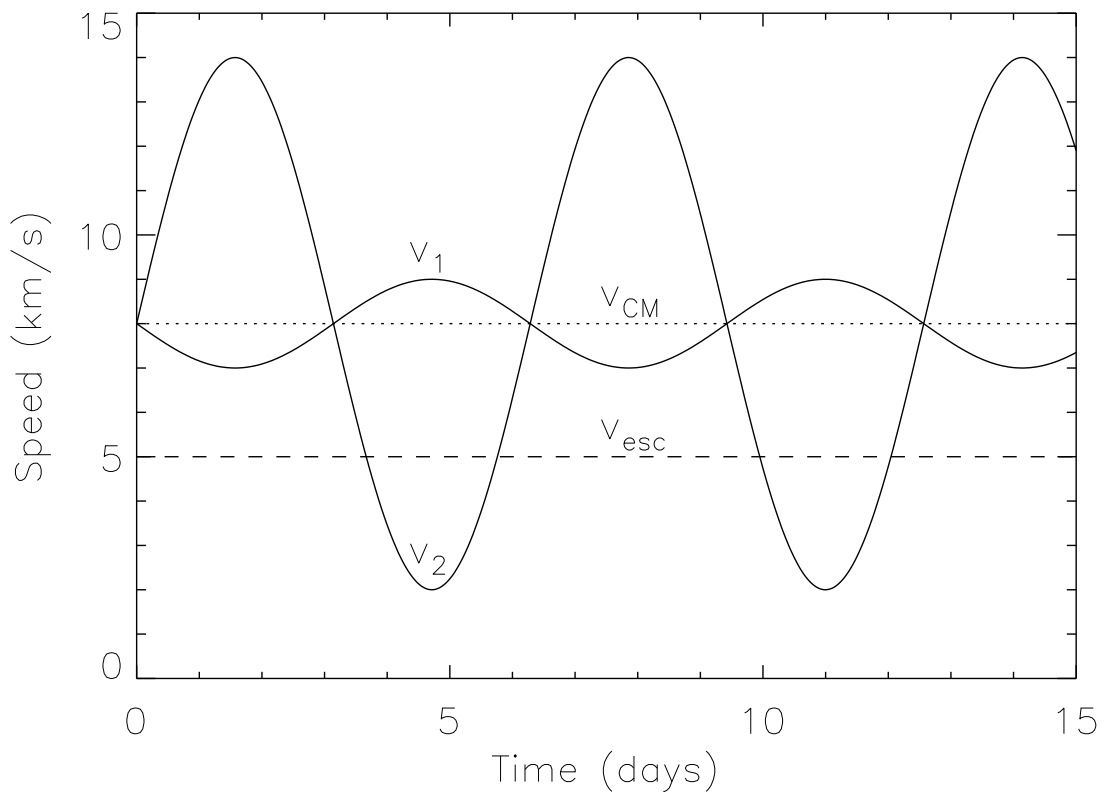}
 \caption{The speeds of unequal-mass binary components ($v_1$ and
  $v_2$) and the center-of-mass speed ($v_{CM}$) relative to Jupiter,
  where $m_1 > m_2$.  Since the binary center of mass approaches the
  planet along a hyperbolic trajectory, it is always traveling faster
  than the escape speed ($v_{esc}$).  In this example, the smaller
  component's speed dips below the escape speed for a portion of its
  orbit.  If the binary is disrupted during this interval, the smaller
  component will be captured by the planet.}
\label{sinecurves}
\end{figure}

As a result of the impulse approximation, we also assume the orbits of
the now-separated components are dictated by their speeds upon
disruption.  The speed change of one component ($\Delta v_{1}$) is
approximately equal to its orbital speed around the CM:

\begin{equation}
\label{deltav}
\Delta v_{1} \approx \pm \frac{m_{2}}{m_{1} + m_{2}} \left(\frac{G (m_{1} +
  m_{2})}{a_{B}}\right)^{1/2} ,
\end{equation}

\noindent where $G$ is the gravitational constant.  If the speed of
either component is below the escape speed ($v_{esc}$) when the binary
is split, that component will be captured.  This is most efficient if
the incoming $v_{CM}$ is only slightly faster than the value needed
for escape.  (See Fig.~\ref{sinecurves}.)

The separation of the binary ($r_B$ = 2$a_B$, for equal-mass pairs on
circular orbits) plays a key role in determining whether a given
encounter will result in a capture.  From Eq.~\ref{deltav}, we can see
that a smaller separation imparts a higher speed change upon
disruption, increasing the probability of capture.  However, the
separation must be large enough that the binary can actually be
disrupted.  Equation~\ref{rtd} indicates that, not surprisingly, a
large separation makes the binary easier to split.  The separation
that optimizes capture, then, is one just wide enough that the binary
is disrupted.  In addition, the tidal radius is important: the speed
change needed for capture ($v_{CM} - v_{esc}$, the difference between
the two horizontal lines in Fig~\ref{sinecurves}) decreases for
smaller $r_{td}$.  Thus deeper encounters are more likely to lead to
captures.

In much of the current work, we consider the simplified case where
Jupiter orbits the Sun along a circle.  In this case, the Jacobi
constant ($C_{J}$) for the planet-Sun-interloper three-body problem is
a very useful predictor of the interloper's potential for capture,
taking on the role of $v_\infty$ from the two-body approximation.
Although our model contains four bodies, and the Jacobi constant is a
three-body construct, it is an excellent approximation to consider the
CM of the binary as one body moving in the Sun-Jupiter system up until
the point of disruption.  The gravitational energy between the binary
components is negligible after they separate.  Thus after disruption,
we essentially have two separate three-body problems, one for each
binary component, and we can make use of the Jacobi constant
throughout the entire simulation.

\begin{figure}[htp]
\plotone{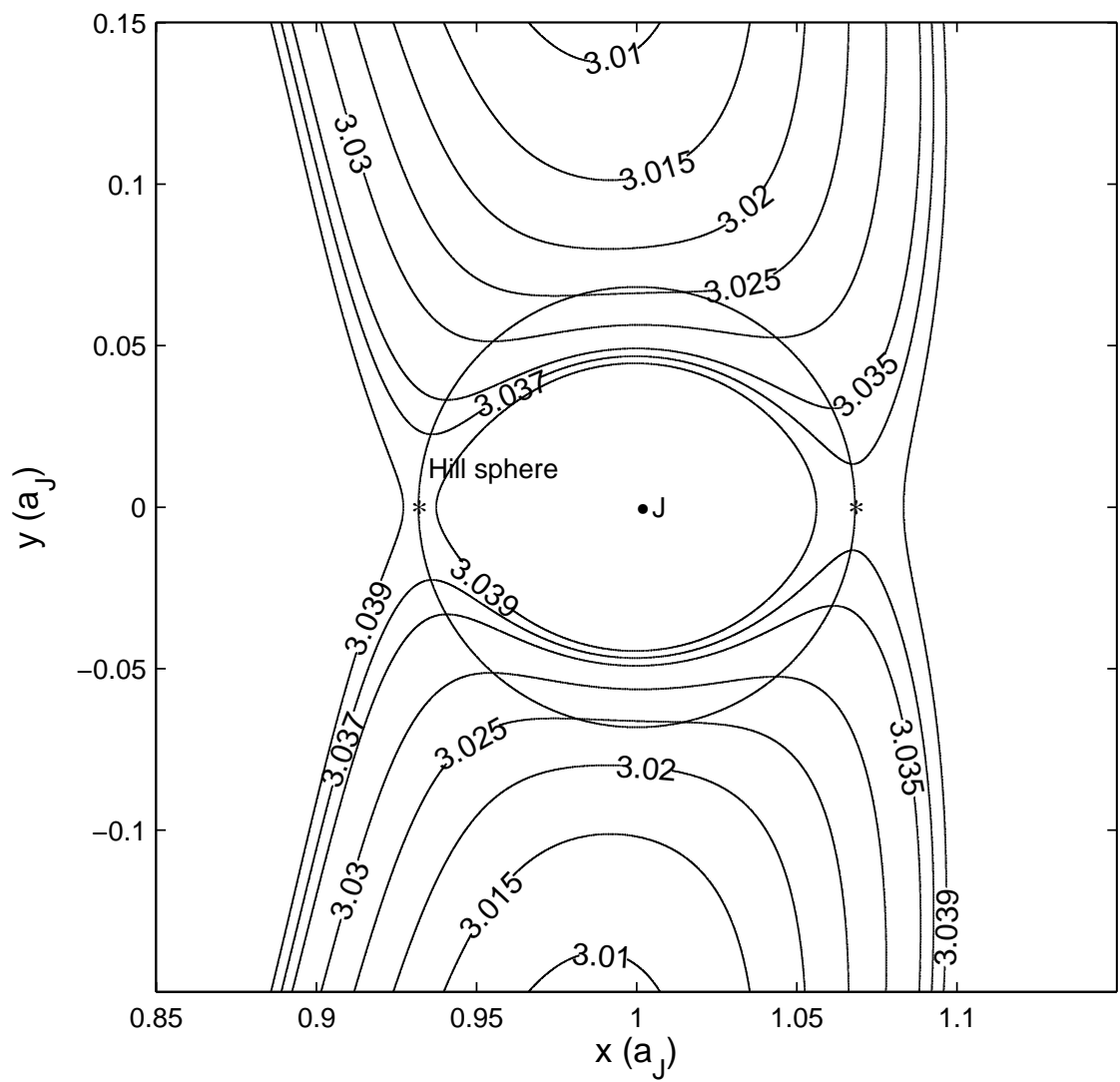}
 \caption{The Hill sphere of Jupiter (central dot marked 'J') with
  zero-velocity curves (ZVCs) corresponding to the labeled Jacobi
  constant ($C_{J}$) values.  The Sun is to the left at (0,0) and the
  Jupiter-Sun separation ($a_J$) is used as the unit of distance.  The
  two asterisks are located at the planet's $L_1$ and $L_2$ Lagrange
  points.  A body near Jupiter with $C_{J} \gtrsim$ 3.0387 (e.g., the
  curve for $C_J$ = 3.039) has ZVCs enclosed around the planet, and it
  would be unable to escape.}
\label{ZVCs}
\end{figure}

If $C_{J} \geq C_{J,crit}$, the critical value for capture, bodies in
the vicinity of the planet are bound by so-called zero-velocity curves
(ZVCs) that enclose Jupiter and constrain particle motions
(Fig.~\ref{ZVCs}).  For Jacobi constants lower than $C_{J,crit}$ (i.e.,
higher energies), one large zero-velocity curve surrounds both Jupiter
and the Sun and bodies can enter and exit Jupiter's Hill sphere
freely.  The critical Jacobi constant represents the boundary between
these possibilities.  Murray \& Dermott (1999) give its value:
$C_{J,crit} \approx 3 + 3^{4/3}\mu^{2/3} - 10\mu/3$, where $\mu$ =
$\frac{M_{P}}{M_{\sun}+M_{P}}$ and $M_{\sun}$ and $M_{P}$ are the
masses of the Sun and the planet, respectively.  Here we use
dimensionless units in which G, the Jupiter-Sun distance, and the sum
of the solar and jovian masses are equal to 1.  For the Jupiter-Sun
system, which is the focus of the current paper, $\mu$ = 9.53 $\times
10^{-4}$ and $C_{J,crit} \approx$ 3.0387.

\begin{figure}[htp]
\plotone{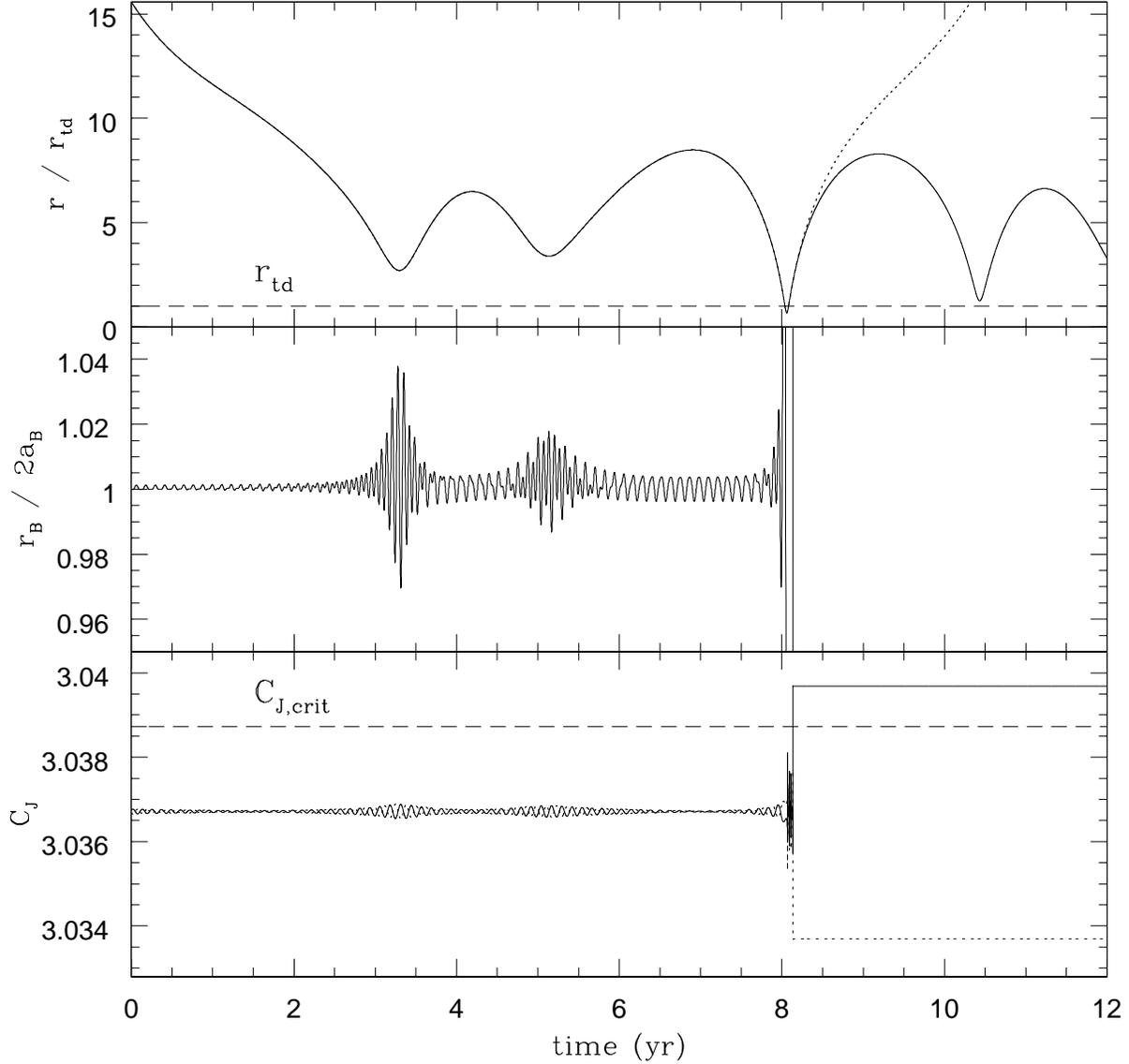}
\caption{Distance from the binary components (each 65-km in radius) to
  Jupiter ($r$) is plotted in units of the tidal disruption radius
  ($r_{td} \approx$ 70 $R_{J}$) in the top panel; the middle panel
  shows the binary's separation ($r_B$) over time, in units of its
  initial separation (2$a_B$); and the bottom panel displays the
  Jacobi constant of each component.  One component is plotted with a
  solid line and the other with a dotted line throughout the figure.
  The dashed line in the top panel indicates the tidal disruption
  radius.  In the bottom panel, the dashed line represents the
  critical Jacobi constant.}
\label{cj3panel}
\end{figure}

Figure~\ref{cj3panel} illustrates a typical capture involving Jupiter.
In the bottom panel, the Jacobi constant of the binary pair prior to
its split is lower than the critical value, meaning that initially,
the binary has too much energy to be bound.  The oscillations in the
bodies' pre-disruption $C_J$ are due to gravitational interactions
between the binary components.  At the time of disruption (t $\approx$
8 yr), one component sharply gains energy ($C_{J}$ decreases), while
the other component experiences a corresponding energy loss ($C_{J}$
increases).  In this example, one component's final $C_{J}$ is higher
than the critical value, signifying that it is permanently bound to
Jupiter.  Though the Jacobi constant is very valuable when considering
a circularly-orbiting Jupiter, a disadvantage is that it cannot be
extended to cases with non-zero eccentricity.  In this paper, we make
the simplifying assumption that $e_{J} = 0$ (rather than the true
value of $\sim$0.048) in order to better elucidate important physics
of the problem.

\begin{figure}[htp]
\plotone{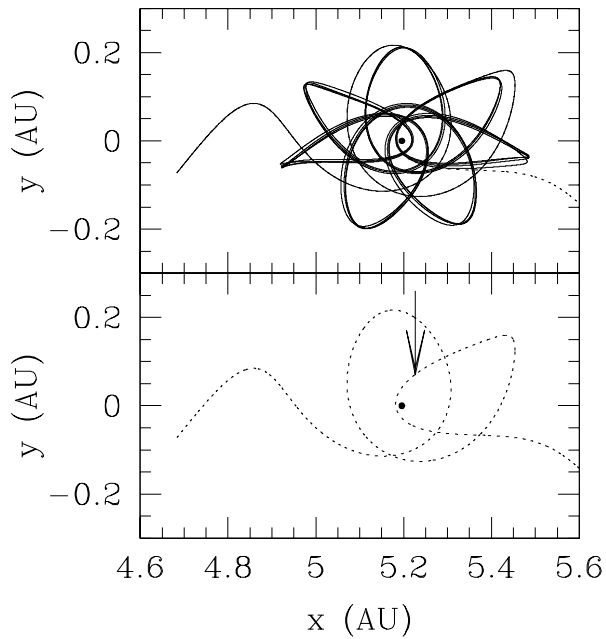}
\caption{The orbits of the binary components discussed in
  Fig.~\ref{cj3panel}.  Jupiter is the dot near the center of each
  panel, and the Sun is located to the left at (0,0).  The components
  are plotted together in the top panel, with solid/dotted lines
  corresponding to those in Fig.~\ref{cj3panel}.  The objects are
  bound to each other as they approach Jupiter from the left side of
  each panel.  A close approach to the planet tidally disrupts the
  binary, sending one component (dotted line) out of the system and
  causing the other component (solid line) to be permanently captured
  by the planet.  The bottom panel contains only the orbit of the
  component that escapes, with an arrow marking the location where the
  binary's separation first exceeds 102\% of its original value.
  About equidistant on the other side of the planet, the binary's
  separation is more than twice its initial value.}
\label{exampleorbit}
\end{figure}

In Fig.~\ref{exampleorbit}, we plot the orbits of the binary
components shown in Fig.~\ref{cj3panel}.  Low-velocity orbits like
these are characterized by multiple close passes by the planet
(cf. Hamilton \& Burns, 1991).  The separation is disturbed by the
strong tidal force during each of these passes, but the binary splits
only after it comes within the tidal disruption radius (see top and
middle panels of Fig.~\ref{cj3panel}).

The binary capture mechanism is most effective at producing permanent
or long-lived captures if i) the mutual orbital speed of the binary is
high, and/or ii) the encounter speed is low.  Agnor \& Hamilton's
(2006a) work examined Neptune's moon Triton, which is somewhat of a
special case because it fulfills both of these criteria -- its size
means that its orbital speed around a close companion would be high,
and typical encounter speeds at Neptune in the early stages of planet
formation are relatively low.
 
The direct three-body capture mechanism is much less effective for
most other irregular satellites which are $\sim$100 km or smaller in
radius.  Furthermore, because of Jupiter and Saturn's sizes and
proximity to the Sun, encounter speeds at the semi-major axes of the
gas giants' irregular satellites are relatively fast, $v_{CM} \approx$
3 km/s.  To produce a large enough energy change for capturing
directly to the current satellites' locations, binary components must
be orbiting each other at speeds comparable to their encounter speeds.
This would require binary companions of order Mars- or Earth-sized
(Agnor \& Hamilton, 2006b) -- an uncommon occurrence even in the early
Solar System.

Accordingly, in this work, we relax the requirement that moons are
captured directly to their present orbits.  In the example discussed
above (see Fig.~\ref{exampleorbit}), the final orbit of the captured
satellite extends almost to the Hill radius ($r_{H}$), whereas the
actual satellites at Jupiter are significantly more tightly bound.  We
investigate the idea that the objects were first captured to these
distant orbits, and a subsequent period of orbital evolution (e.g., by
weak gas drag) led them to their current configurations.

The post-capture evolution is a key component in our model because it
allows for capture from small binary pairs, even though they deliver
satellites to very distant orbits.  Binaries with primaries of order
100 km were certainly much more numerous than those with planet-sized
primaries, even in the early Solar System.  Models that rely on gas
drag for capture (e.g. Pollack \etall, 1979) require both i) dense gas
(to enable capture) and ii) rapid dispersal (to prevent satellite loss
to the planet).  By contrast, our model requires no gas for capture
and puts only weak constraints on gas required for orbital evolution.
In particular, we require only that the product of the gas density and
its residence time around Jupiter be large enough that the requisite
amount of evolution can occur.

\section{Numerical model}

The goal of this work is to characterize the overall effect of
binaries on the probability of capturing bodies on planet-crossing
paths.  We focus primarily on captures at Jupiter, which has the most
irregular satellites and has many sources of small bodies nearby.
Also, as discussed above, capture at Jupiter has shown to be
difficult, especially because of its large size and fast encounter
speeds for approaching bodies.  Thus these simulations provide the
most stringent test of our model.

Our integrations include the Sun, Jupiter, and a binary or single
object, in a planet-centered frame.  In order to examine binaries'
effectiveness at producing long-lived captures, we compare them to
captures of single-body interlopers.  While only tidally disrupted
binaries can be captured permanently, unbound single bodies can remain
near the planet for long periods of time (e.g. Astakhov \etall, 2003).
We define a 'capture' to be a body that remains near Jupiter for 1,000
years, the duration of each simulation.  Furthermore, we define
'binary capture' to mean capture of one or both of the bodies that
originated together as binary components.  Note that these definitions
encompass both permanent (energetically bound) and long-lived
temporary captures.  With captures of single bodies as a baseline, we
are able to measure the enhancement due to binaries.

Our simulations are performed with HNBody, a hierarchical {\it N}-body
integration package, and HNDrag, a companion code for applying
non-gravitational forces to the particles and for detecting close
approaches (Rauch \& Hamilton, 2002).  For most of this work, we use
only the close-approach detecting capabilities of HNDrag and include
only gravitational forces.  We use HNBody's Bulirsch-Stoer integrator
with a specified accuracy of one part in $10^{14}$.  The
adaptive-stepsize Bulirsch-Stoer integrator is much more efficient
than a symplectic integrator here because while a small stepsize is
needed initially to resolve the orbital motion around the binary CM,
it can be greatly increased after disruption.

We ran about 200 sets of three- or four-body simulations examining a
range of Jacobi constants for each interloper (2.95-3.037), as well as
varying the binaries' radii (65-, 100-, and 125-km), and separations
(1-1000 body radii).  For each set of parameters, we generated 10,000
binaries or single objects, for a total of $\sim$2$\times10^{6}$
simulations, each following the bodies for 1,000 years.  We started
all of the interlopers of a given set at the same distance from the
planet, ranging from 1.0 - 1.4 $r_H$.  (Section~\ref{startingdistance}
contains a discussion of the effects of starting distances on capture
statistics.)  The choice of Jacobi constant and starting distance
constrains the possible initial positions of the binary CM.
Fig.~\ref{ZVCs} shows Jupiter's Hill sphere overplotted with ZVCs
corresponding to the Jacobi constants that we studied.  Bodies are
energetically unable to cross their zero-velocity surfaces, and thus
starting with, say, $C_{J}$ = 3.037 at 1.0 $r_{H}$ from the planet
restricts the body's initial position to the two small 'endcaps' of
the Hill sphere along the Jupiter-Sun line.  For smaller $C_{J}$,
these allowed areas are larger and finally encompass the entire Hill
sphere for $C_{J} \lesssim $3.025.  The bodies' initial speeds are
also constrained by the specified Jacobi constant, and we choose the
velocity to point in random inward directions.

For simplicity, we set the binary components to orbit each other on
circles and the binary angular momentum to be perpendicular to
Jupiter's equatorial plane.  We ran each initial condition with five
different binary orbital phases (equally-spaced mean anomalies) and
averaged all capture statistics over the five phases.  Throughout the
simulations, we monitored the bodies, weeding out very close
approaches between any two objects and noting each body's close
approaches to Jupiter.  (Collisions between binary members do occur,
but these are rare and of limited interest, since the merged object
simply behaves as a single interloper with the same CM speed.)  To
shorten the computational time required, we stopped integrations in
which all of the incoming objects traveled further than 2-5 Hill radii
from Jupiter, depending on the bodies' starting distance from the
planet.

\section{Results}

\subsection{Relationship between inclination and $C_{J}$}

\label{incl}

\begin{figure}[htp]
\plotone{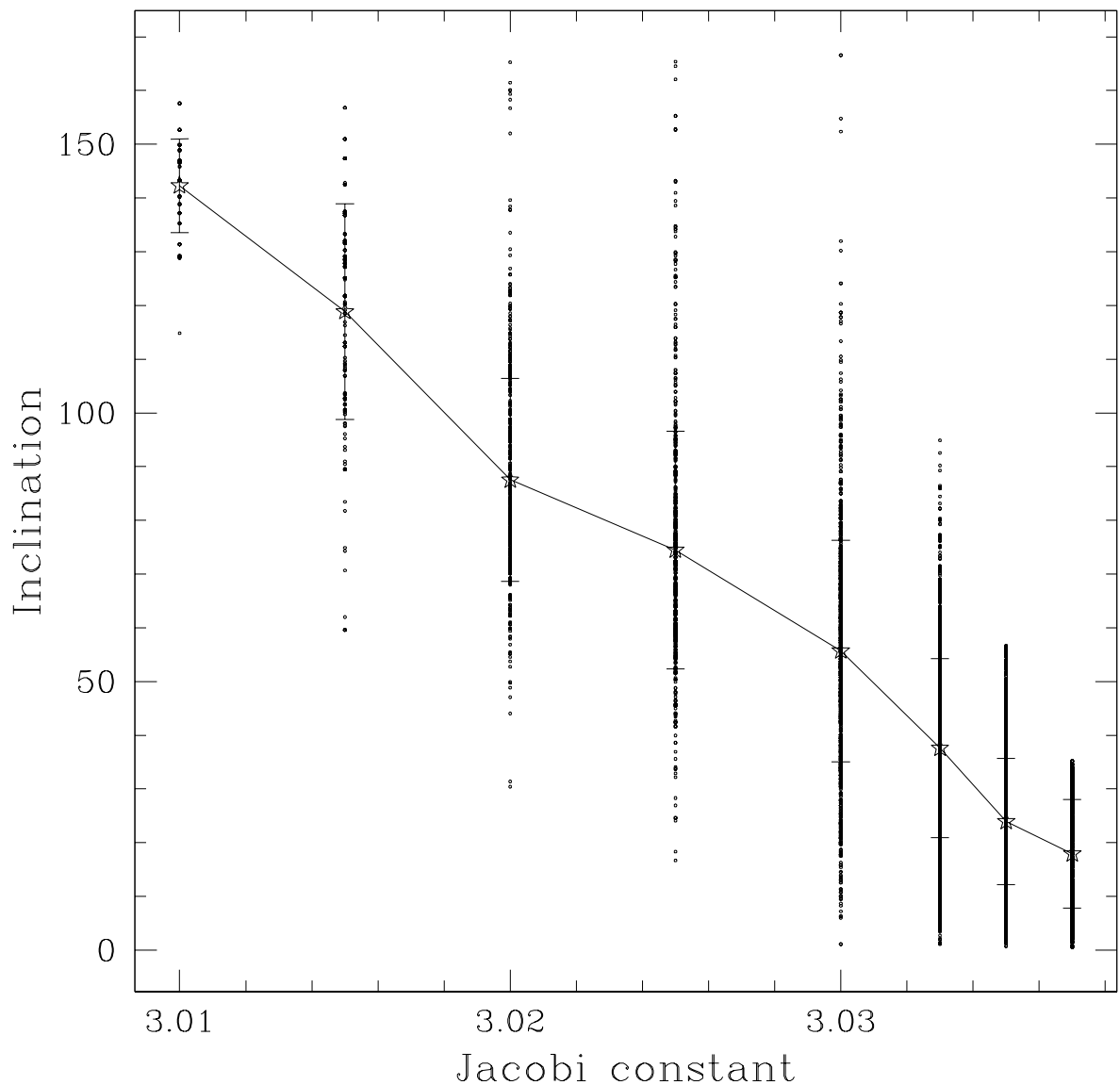}
\caption{Inclination of 100-km captured binary components as a
  function of initial Jacobi constant, plotted for the bodies' first
  close approaches.  Similar plots show that the relationship holds
  for all close approaches of binaries or single bodies, captured or
  not.  Also plotted is a line connecting the mean at each Jacobi
  constant (marked by the stars) as well as 1-$\sigma$ error bars.
  See Section~\ref{incl} for discussion on the $C_{J}$-inclination
  relationship.}
\label{inclCJ}
\end{figure}

In our simulations, we find that the inclination of the approach
trajectory is correlated with the initial Jacobi constant, which is
helpful in providing physical intuition for the meaning of $C_{J}$.
This correlation was first noticed numerically by Astakhov \etal
(2003); here we confirm their finding numerically and provide an
analytical explanation.  Fig.~\ref{inclCJ} displays this
$C_J$-inclination relationship.  We see a clear correlation of $C_J$
with mean inclination: lower Jacobi constants are indicative of
retrograde orbits, while prograde orbits have larger $C_J$.  For
clarity, the plot shows only a representative population of bodies:
100-km binary members that result in a capture, with inclinations
calculated at the closest approach of each body's first pass by
Jupiter.  However, the relationship holds for all close approaches of
binaries or single objects, captured or not.  This plot provokes two
main questions: what is the physical cause of this trend and why is
there such high scatter in inclination at a given Jacobi constant?  We
address the question of scatter first.

One complexity in making this plot is that all orbital elements
including inclination are poorly defined at large distances from
Jupiter, as solar tides are comparable to Jupiter's gravity at the
Hill sphere.  Accordingly, we were careful to calculate the
inclinations only at orbital pericenter where solar tides are weakest
so that inclination is always well defined.  Poorly-defined orbital
elements, therefore, are not the source of the scatter.  Furthermore,
the variations look nearly the same when we plot single objects rather
than binaries, which is expected since disrupting the binary results
in an energy change that only slightly alters $C_J$ (e.g.,
Fig.~\ref{cj3panel}).  Finally, the scatter is present even when we
consider one individual object's multiple pericenter passages rather
than those of an ensemble of objects.  Thus the spread in inclination
is real and is due to the response of a single captured object to the
solar tidal force.

The scatter in inclination as well as the inclination-$C_{J}$ trend
can be understood analytically by writing the Jacobi contant in terms
of planetocentric orbital elements rather than the heliocentric
orbital elements used in deriving the standard Tisserand constant
(Murray \& Dermott, 1999).  We begin with the planet-centered
'generalized Tisserand constant' derived in Hamilton \& Krivov, 1997
(their Eq. 4) and neglect the solar tidal term since it is complicated
and unimportant at pericenter where we measure inclination.  We then
non-dimensionalize the equation as described in
Section~\ref{3bodycapture}, and finally, to conform to standard usage
(Murray \& Dermott, 1999), we add the constant 3 to the final result
and find:

\begin{equation}
\label{cjinclrel}
{C_{J}}' = 3 + \frac{3^{1/3}\mu^{2/3}}{\bar{a}}\left[1 + 2\left(\frac{\bar{a}^{3}(1-e^{2})}{3}\right)^{1/2}\cos(i)\right],
\end{equation}

\noindent where $\mu$ is the mass ratio as defined above;
$\bar{a} = a/r_{H}$; and $a$, $e$, and $i$ are the captured satellite's
semi-major axis, eccentricity, and inclination, respectively.  Because
we have neglected the solar tidal term and used planetocentric orbital
elements, this expression is valid only near the planet where solar
perturbations are weak.

For close orbits of the planet ($\bar{a} << r_H$), orbit-averaging the
effects of the tidal force shows that the semimajor axis $\bar{a}$
remains constant.  Accordingly, Eq.~\ref{cjinclrel} leads directly to
the Kozai constant, $K = \sqrt{1 - e^{2}}\cos(i)$.  This constant
explains the coupled oscillations in eccentricity and inclination that
characterize the Kozai resonance.  If $K$ were precisely conserved,
orbits would not be able to switch between $i < 90^\circ$ (which have
$K > 0$) and $i > 90^\circ$ (which have $K < 0$).  We do, however, see
such prograde-to-retrograde transfers in Fig.~\ref{inclCJ}, which
indicates that, as expected, $K$ (and therefore $\bar{a}$) is not
constant for our distant orbits (see, e.g., Hamilton \& Burns, 1991).
In addition, at apocenter where the solar tidal force is strongest,
the orbital elements themselves are poorly defined and
Eq.~\ref{cjinclrel} is only approximate.  Thus between each pericenter
passage, the orbital elements (including inclination) are scrambled by
the solar tidal force leading to dispersion like that seen in
Fig.~\ref{inclCJ}.

The trend observed in Fig.~\ref{inclCJ}, decreasing inclination for
increasing Jacobi constant, is neatly explained by
Eq.~\ref{cjinclrel}. Testing the equation quantitatively, we estimate
$\bar{a}$ = 0.5 and $e$ = 0.7 for a typical orbit of a body captured
at Jupiter (e.g., Fig.~\ref{exampleorbit}). A purely prograde orbit
($i = 0$) gives ${C_{J}}'$ = 3.036, while a purely retrograde orbit
($i = 180^\circ$) gives ${C_{J}}'$ = 3.020.  These values roughly
correspond to the range of Jacobi constants seen in Fig.~\ref{inclCJ},
despite the rather large approximations that we have made.  The
inclination-$C_J$ correlation is strong enough that we will often use
the term prograde to refer to orbits with $C_{J}\sim3.03$, and
retrograde to mean $C_{J}\sim3.01$.

\subsection{Modes of capture}

\label{modes}

\begin{figure}[htp]
\plotone{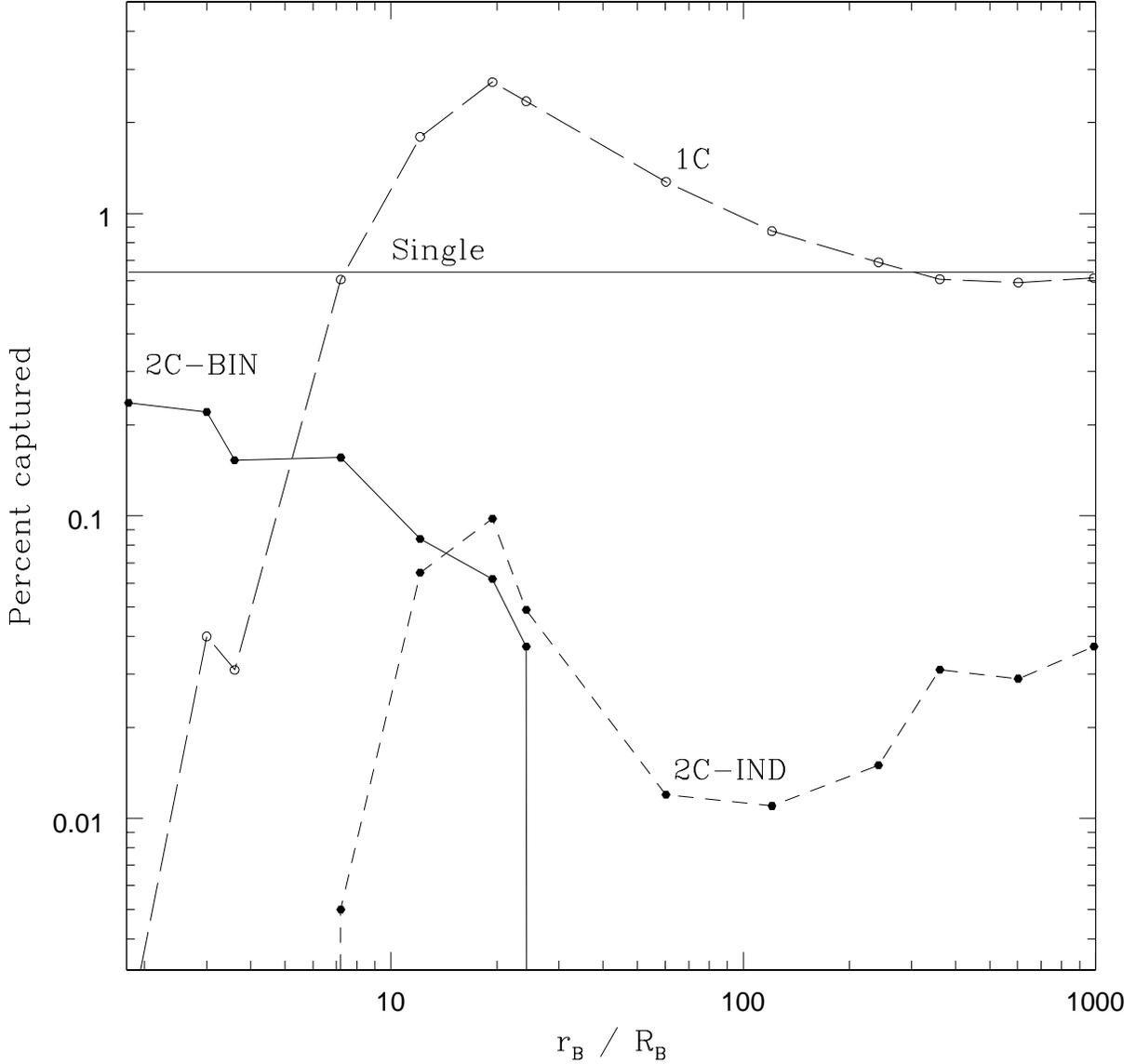}
\caption{Three modes of capture for integrations of $R_B$ = 65-km
  binary pairs with $C_{J}$ = 3.037, as a function of the separation
  of the binary ($r_B$): one component captures ('1C'), the binary
  splits and both capture independently ('2C-IND'), or the binary
  remains bound and captures as a pair ('2C-BIN').  We plot the
  capture percentage for objects (rather than binaries) to facilitate
  comparison with single bodies (upper solid line).  Recall our
  working definition of capture to mean bodies still orbiting the
  planet after 1,000 years.  Note that the 2C-BIN curve approaches the
  value for singles at small separations while the sum of the 1C and
  2C-IND captures rates approaches the same value for large
  separations.}
\label{modes3.037}
\end{figure}

For each binary-planet encounter, there are four possible outcomes:
(1) neither component captures (hereafter known as '0C'), (2) one
component captures ('1C'), (3) both components capture together as an
intact binary without splitting apart ('2C-BIN'), or (4) the binary is
disrupted and both components capture individually ('2C-IND').  The
frequency of each type of outcome depends on the characteristics of
the binary.  Fig.~\ref{modes3.037} shows the outcomes that result in a
capture (i.e., 1C, 2C-BIN, and 2C-IND) for 65-km binary pairs with
initial $C_{J}$ = 3.037, as a function of the initial separation,
$r_B$, of the binary.  The separation can be altered significantly
prior to disruption during close approaches to the planet.  The number
of captures for a set of single objects is also plotted for
comparison; these must be temporary captures since there is no energy
loss.

The rate of 2C-BIN captures is largest when the separation is small
(and thus the components are tightly bound to each other).  For
small-enough separations, the tidal disruption radius is so close to
the planet that very few binary orbits cross it (see Eq.~\ref{rtd}).
Here, most of the binaries remain intact, and the 2C-BIN rate nearly
matches that of single objects.  When we increase the binary
separation, more binaries are split, and the 2C-BIN capture percentage
monotonically drops to zero, as expected.

Disrupting binaries leads to more possibilities for capture of
individual objects.  Accordingly, as the separation increases, the 1C
and 2C-IND capture rates rises from zero.  For the 1C population of
65-km objects with $C_{J}$ = 3.037, there is a peak in capture
efficiency of $\sim$5 times that of a single body at a separation of
$\sim$20 body radii ($R_B$).  This separation represents the optimal
balance between disrupting a high percentage of the binaries and
delivering the most energy upon disruption.  The optimum separation
varies depending on the mass of the binary.  At larger separations,
the binding energy decreases, leading to smaller energy kicks and a
diminished capture rate.

The 2C-IND percentage has a peak at the same separation as the 1C
group.  These binaries likely split during orbital phases where the
energy is distributed almost equally between the components.  The
number of 2C-IND captures is never more than a few percent of the 1C
captures, but the two populations peak at $r_B \sim$ 20 $R_B$ for the
same reasons.  More widely separated binaries are disrupted with a
smaller energy change.  Because of this, the two components are more
likely to have similar energies and post-disruption fates, causing an
increase in 2C-IND captures at larger separations.

Unlike the case for 1C and 2C-IND capture where energy is lost and
capture can be long-lived or even permanent (as in
Fig.~\ref{cj3panel}), capture of singles or intact binaries (2C-BIN)
is necessarily temporary.  This could be an advantage for 1C captures,
which have more stable, lower-energy initial capture orbits.  The
details of the final comparative satellite yields depend on the
subsequent orbital evolution, which is determined by the gas present
at the time of capture and its dissipation timescale.

\subsection{Effects of binary mass and orbital separation}

Having explored the physical meaning of $C_J$ and the possible types
of captures, we now discuss the results of the numerical simulations.
In this section, we consider cases of equal-mass binaries encountering
Jupiter with the planet on a circular orbit, and we examine the
effects of the bodies' masses, binary separations, and initial Jacobi
constants.  We performed integrations over a range of Jacobi
constants: 2.95 $\leq C_{J} \leq$ 3.037, where $C_{J} \approx$ 3.0387
is the critical value above which transfer orbits between Jupiter and
the Sun are impossible (see Fig.~\ref{ZVCs}).  For $C_{J} \leq$ 2.99,
no captures resulted for any of the parameters we tested, although
capture at these low Jacobi constants could certainly occur for
larger-mass binaries.  For now, we consider the fate of bodies started
from the Hill sphere (following Astakhov \etall, 2003); in
Section~\ref{startingdistance}, we discuss the importance of
alternative starting distances.

\begin{figure}[htp]
\plotone{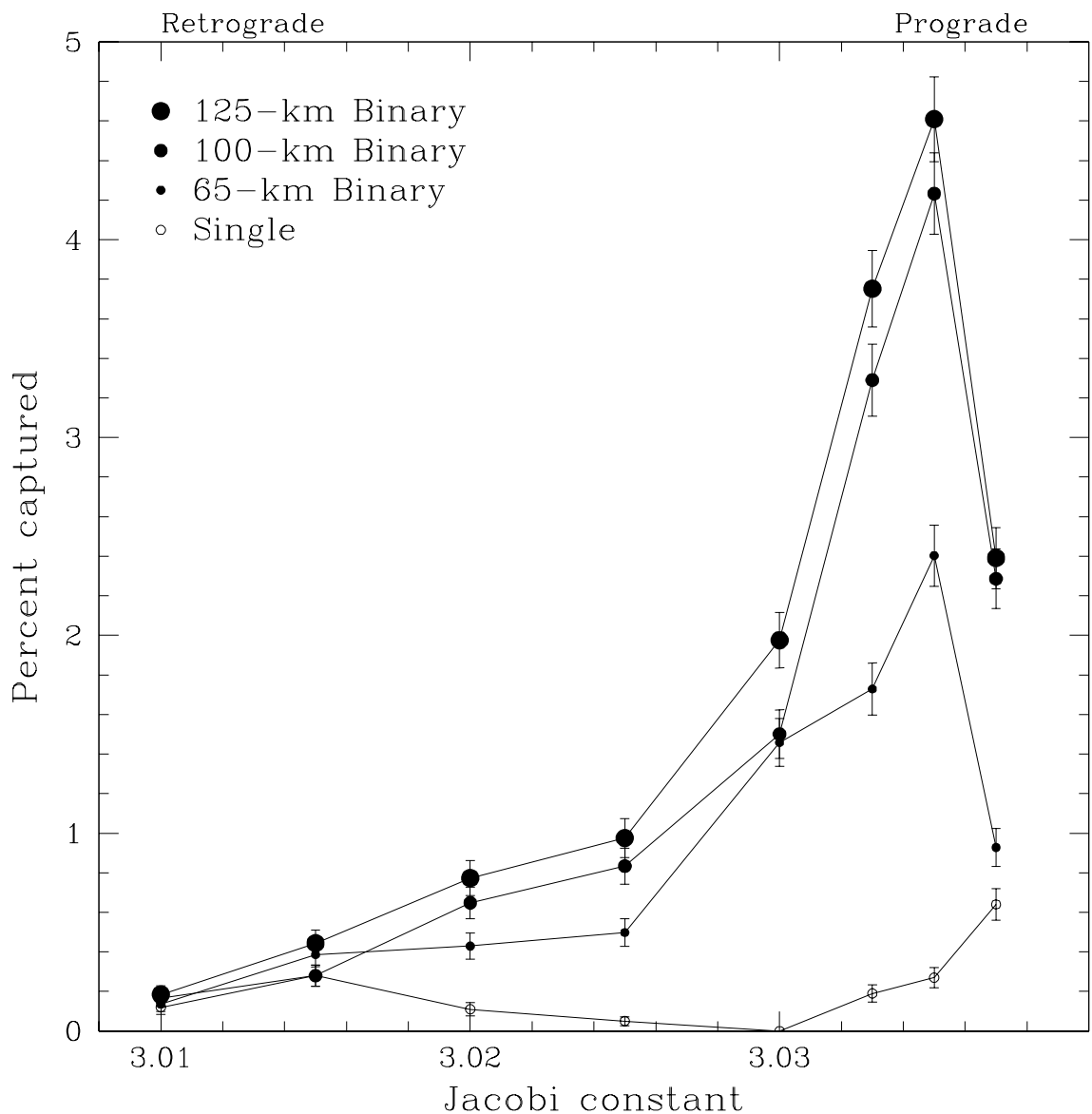}
\caption{Capture percentages for binaries of different masses compared
  to single objects.  All bodies were started on Jupiter's Hill
  sphere.  For each binary mass, a single separation was used over the
  range of initial Jacobi constants: 471 km = 7.25 binary radii
  ($R_B$) for the 65-km binaries, 1225 km = 12.25 $R_B$ for the 100-km
  set, and 1512 km = 12.10 $R_B$ for 125-km binaries.  These
  separations give near maximum capture rates for the majority of the
  Jacobi constants tested, with the exception of the highest $C_J$,
  where the optimum separation is closer to 20 binary radii.}
\label{1rH}
\end{figure}

We examined masses corresponding to pairs of objects each with radii
65-km, 100-km, and 125-km (assuming a density of $\sim$2 $g/cm^{3}$).
Fig.~\ref{1rH} displays the results of these mass studies.  We see
that capture rates increase for higher masses: the 125-km capture rate
is slightly higher than the 100-km rate throughout the range of Jacobi
constants, and they differ most significantly from 65-km binary pairs
for $C_{J} >$ 3.03.

For single objects, mass has no effect on capture probability, but for
binaries, larger total mass leads to more rapid orbital speeds and a
higher speed change upon breakup.  This can be seen by eliminating
$a_B$ from Eqs.~\ref{rtd} and~\ref{deltav}, with mass ratio
($\frac{m_{2}}{m_{1} + m_{2}}$) and tidal distance ($r_{td}$) held
constant; the result is $\Delta v_{1} \sim (m_{1} + m_{2})^{1/3}$.
Accordingly, larger masses generally lead to increased capture rates.

Another important result is that capture rates from binaries are
extremely sensitive to the binary's separation, $r_B$ = 2$a_B$.  For
each of the masses we examined, we determined the optimum separation
of the binary required to achieve the maximum capture probability.  We
used Eqs.~\ref{rtd} and~\ref{deltav} to guide our separation
selection, and we integrated each point on Fig.~\ref{1rH} with several
different binary separations to determine the optimal value.

In Fig.~\ref{1rH}, we have plotted statistics using a single
separation for each mass over the range of Jacobi constants.  For most
of the Jacobi constants studied for a given mass, the optimal
separations are very similar, $\sim$10 $R_B$.  An important exception
is for the highest $C_{J}$ value tested, 3.037 (recall that this
corresponds to mostly prograde encounters -- see Fig.~\ref{inclCJ}),
which had maximum captures at a larger separation ($\sim$20 $R_B$)
than the typical optimal value.  Because we have not plotted this
optimal value in Fig.~\ref{1rH}, the curve declines sharply at $C_{J}$
= 3.037. As is clear from Section~\ref{modes}, optimizing the
separation makes a significant difference in capture rate, especially
for binaries whose $C_{J}$ values are close to the capture threshold.
Changing from $\sim$10 $R_B$ to the optimal 20 $R_B$ for $C_{J}$ =
3.037 increases the capture percentage from $\sim$2$\%$ to
$\sim$10$\%$ (100-125 km objects) and $\sim$1$\%$ to $\sim$3$\%$
(65-km, see Fig.~\ref{modes3.037}).

Binary capture rates also appear to depend strongly on initial Jacobi
constant.  It is tempting to compare capture efficiencies for low and
high Jacobi constants (retrograde and prograde orbits).  A direct
comparison of these rates, however, cannot be made for reasons that
will become apparent in the next section.  We can, however, compare
the binary statistics to those of single objects.  The bodies that
originate in binaries capture with similar rates as the single objects
below $C_{J} \approx$ 3.015, but as $C_{J}$ increases, the effects of
the binaries become more visible, rising by an order of magnitude in
efficiency at delivering objects to Jupiter.  Part of the reason for
this is probably that the retrograde binaries are harder to split than
progrades because of their orientation to the planet.  Another
explanation, particularly for the highest initial Jacobi constants, is
that these encounters are close to the critial energy barrier for
capture, and so the energy change from disruption of the binary is
more likely to result in capture.

\subsection{Effects of starting distance}

\label{startingdistance}

\subsubsection{Contamination from bound retrograde orbits}

\label{subretr}

Thus far we have discussed results from initial conditions that launch
objects from the Hill sphere.  In the course of this study, we
discovered that the choice of starting distance can significantly
affect the capture statistics.  While the Hill sphere is defined as a
rough stability boundary beyond which the Sun's gravitational
influence is stronger than the planet's, in practice, stable
retrograde orbits can extend out to distances slightly beyond the Hill
radius (see, e.g., Hamilton \& Burns, 1991).  In contrast, stable
prograde orbits (e.g., Fig.~\ref{exampleorbit}) are always well within the
Hill sphere.  Starting the integrations with bodies on the Hill
sphere, then, risks starting on an already-stable retrograde orbit.
This causes ambiguity in the capture statistics -- which orbits were
always near Jupiter and which truly came in from infinity?

We demand that true captures originate in heliocentric orbit and
transition to planet-centered orbits, remaining for at least 1,000
years.  To differentiate between these and misleading 'captures' from
bodies that are already orbiting the planet at the beginning of our
integration, we took the single-body capture orbits and integrated the
initial conditions backwards in time for 1,000 years.  We then
separated the captures that came in from infinity and those that were
always present near the planet.  We found that when starting on the
Hill sphere, most of the resulting retrograde captures of single
objects at low $C_{J}$ had always been orbiting the planet and were in
fact not true captures.

\begin{figure}[htp]
\plotone{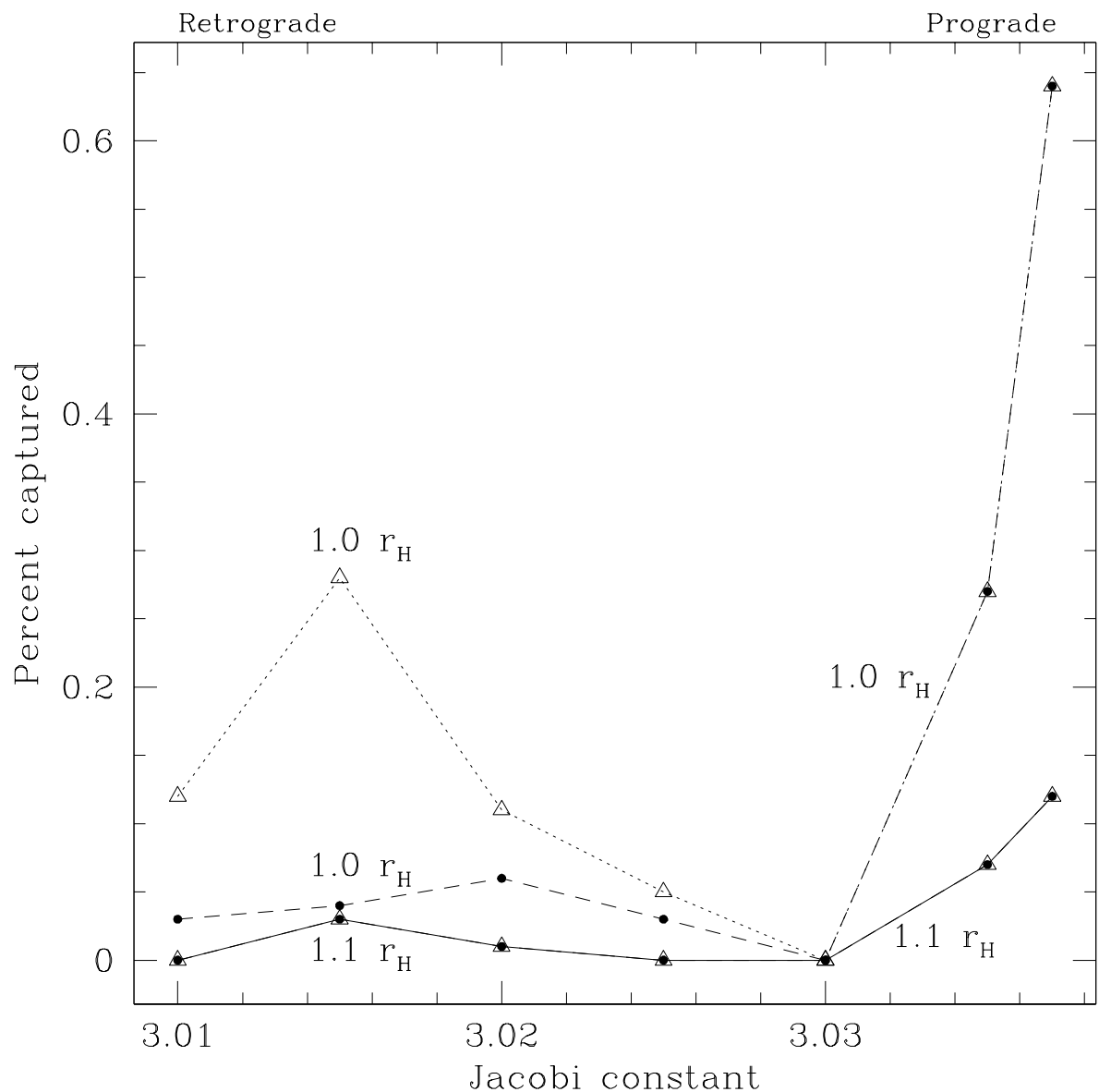}
\caption{Single objects integrated from two launch distances.  True
  captures curves (with points that are solid circles) show only those
  captures that originated far from Jupiter, while curves that show
  true plus contaminated captures (with open triangles) also include
  objects that were stably orbiting the planet at the beginning of the
  integrations.  For launch on the Hill sphere (1.0 $r_{H}$), we see
  that the prograde orbits (high $C_{J}$) are not affected at all,
  while many of the retrograde orbits (low $C_{J}$) are discovered to
  be false.  For launches at 1.1 $r_{H}$, the true and
  true-plus-contaminated curves overlap completely, showing that
  contamination by false captures is effectively zero.}
\label{rHbkwd}
\end{figure}

We examined several other launch distances and determined that
starting from 1.1 Hill radii ($r_{H}$) and beyond eliminates nearly
all contamination from already-stable retrograde orbits.
Figure~\ref{rHbkwd} compares the capture rates for single objects
beginning from 1.0 $r_{H}$ and 1.1 $r_{H}$.  We see that while the
high-$C_{J}$ captures are uncontaminated, launching from 1.0 $r_{H}$
for $C_{J} <$ 3.03 leads to many false captures.  In fact, over
one-third of the 1.0 $r_{H}$ captures are contaminations.  (This
fraction is much smaller for binaries, where there are many more
prograde captures than retrograde.)  When the objects are launched
from 1.1 $r_{H}$, we see no artificial captures at all.  Similar tests
at 1.2 $r_H$ and 1.4 $r_H$ also show no false captures.  Although
stable orbits do exist to these and larger distances (Henon, 1969),
they are apparently extremely rare.

\subsubsection{Scaling to different starting distances}

So, is it safe to compare prograde and retrograde capture statistics
for starting distances beyond 1.1 $r_H$?  Starting outside the Hill
sphere results in a decrease in true captures over the full range of $C_J$
(as seen in Fig.~\ref{rHbkwd}).  This is expected, because fewer of
these bodies experience close approaches with the planet.  However,
prograde captures are more drastically reduced than retrogrades.
Determining the reason for this is critical to answering our question.

At high Jacobi constants ($\gtrsim$ 3.03, which correspond to mostly
prograde orbits), the binary's ZVCs are closed around the planet
except for only a small neck on either side of Jupiter, toward and
away from the Sun (see Fig.~\ref{ZVCs}).  In contrast, lower Jacobi
constants (and thus retrograde orbits) allow for more freedom of
movement in the region between the Sun and Jupiter.  Retrograde orbits
can approach the planet from all directions, while progrades are
limited to the entering via the narrow ZVC necks.  Starting further
than the Hill sphere means that a smaller fraction of the initial
prograde trajectories pass through the zero-velocity necks and
approach Jupiter, resulting in a decrease in captures.  We tested the
hypothesis that the shape of the prograde ZVCs is the primary reason
for the decrease in prograde captures with the following procedure:

\indent 1) Determine the capture rate at 1.2 $r_H$.  

\indent 2) Determine the crossing rate of 1.2 $r_H$ orbits inside 1.1 $r_H$.

\indent 3) Scale the 1.1 $r_H$ capture rate by the crossing rate from
step (2).  

\begin{figure}[htp]
\plotone{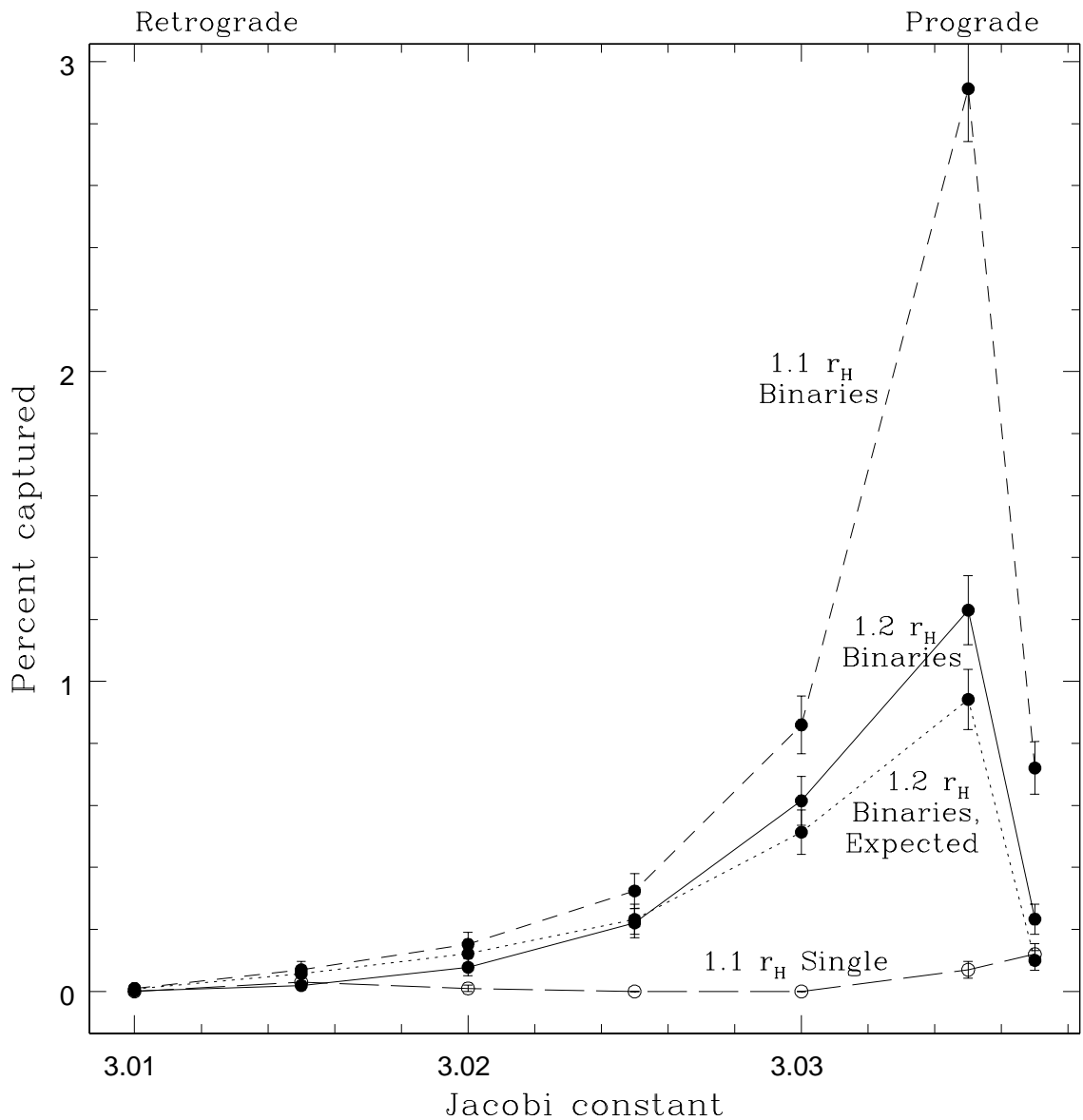}
\caption{Capture rates of 100-km binaries that were launched from 1.1
  $r_{H}$ (upper dashed line), 1.2 $r_{H}$ (solid line), and expected
  capture rates for 1.2 $r_{H}$ (dotted line) calculated by scaling
  the 1.1 $r_{H}$ rates by the percent of 1.2 $r_{H}$ trajectories
  crossing interior to the 1.1 Hill radii.  All are shown with
  1-$\sigma$ error bars.  This scaling equalizes the capture
  percentages between the two launch distances to within 20\%.  For
  reference, the capture rate for single bodies starting from 1.1
  $r_{H}$ is also plotted; compare with Fig.~\ref{1rH}.}
\label{scaling}
\end{figure}

\noindent We use the statistics for 1.1 $r_{H}$ and 1.2 $r_{H}$
because they are free from any contamination from the false retrograde
captures discussed above.  The result of step (3) is what we expect
the 1.2 $r_{H}$ capture percentages would be if the ZVC shapes are the
reason for the decline in prograde captures from 1.1 $r_{H}$ to 1.2 $r_{H}$.
Comparing these scaled capture rates with the actual 1.2 $r_{H}$
percentages (as shown in Fig.~\ref{scaling}), we see that this scaling
accounts for most of the difference between the two capture rates, to
within 20\%.  This indicates that our prediction is valid.

In principle, these results also account for the reduction in true
retrograde captures, but the number statistics are so low for
retrogrades at these launch distances that the actual and scaled
capture percentages are equivalent to within the error.

We find, then, that because of the differing geometries of their ZVCs,
we cannot directly compare prograde and retrograde statistics.  This
is true even when starting beyond the limit for already-stable
retrograde orbits, chiefly because prograde orbits are so sensitive to
their initial distance.  To circumvent this effect, we would have to
start far enough away from the planet that further increasing the
starting distance would result in an equal fractional decrease in
captures for retrogrades and progrades -- in other words, where the
geometry of the prograde ZVCs is no longer the dominant reason for the
decrease.  Our simulations, out to 1.4 $r_{H}$, did not reach this
threshold and were limited by number statistics.  We anticipate that
comparing prograde and retrograde captures would require starting still
further from the planet than our trials, and would demand much larger
initial populations.

\subsubsection{Starting from 1.1 $r_H$ vs. 1.0 $r_H$}

Launching binaries from 1.1 $r_{H}$ rather than 1.0 $r_{H}$ results in
an overall decrease in captures for the reasons described above, but
many of the characteristics of the capture orbits remain relatively
unchanged.  For example, the inclination distribution for the 1.1
$r_{H}$ captures is almost identical to the 1.0 $r_{H}$ population
plotted in Fig.~\ref{inclCJ}.  Also, comparing Fig.~\ref{scaling} with
Fig.~\ref{1rH}, we see that the trend in the capture percentage with
$C_J$ of the 100-km binaries at 1.1 $r_{H}$ and 1.0 $r_{H}$ are
generally similar.

\begin{figure}[htp]
\plotone{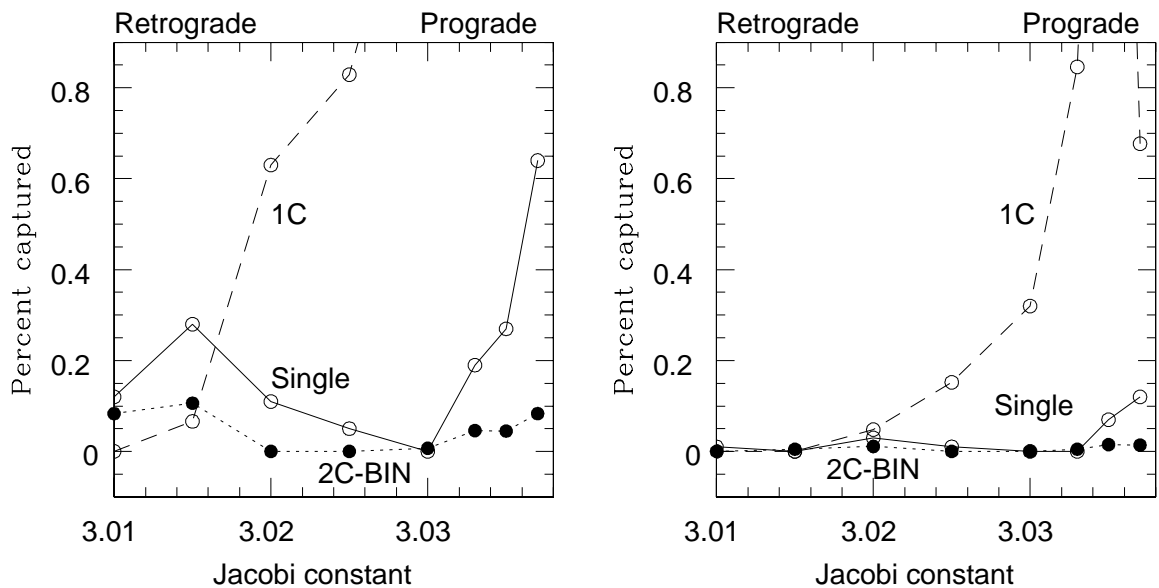}
\caption{Modes of capture vs. Jacobi constant, for integrations of
  100-km binaries starting at 1.0 $r_{H}$ (left panel) and 1.1 $r_{H}$
  (right panel).  The 1C curve peaks are off the top of the plot at
  $\sim$4.2\% for the 1.0 $r_{H}$ runs and $\sim$3.0\% for the 1.1
  $r_{H}$ group.  The 2C-IND capture rates are extremely small for
  both starting distances (e.g., Fig.~\ref{modes3.037}) and, for
  clarity, are not plotted here.  The capture rates of single bodies
  are plotted for comparison.}
\label{modesCJ}
\end{figure}

Figure~\ref{modesCJ} displays the modes of capture vs. Jacobi constant
for 100-km binaries and single objects starting from both 1.0 $r_{H}$
and 1.1 $r_{H}$.  Starting at 1.1 $r_{H}$ results in a depletion of 2C
captures relative to 1C captures.  This is because many of the
retrograde 2C-IND and 2C-BIN captures were likely the artificial
retrogrades which are eliminated when starting from 1.1 $r_{H}$.  The
single captures also appear depleted relative to the other curves, but
in fact, the binaries are diminished by a roughly comparable amount.
Overall, there are few qualitative differences between the 1.0 $r_{H}$
and the 1.1 $r_{H}$ cases.

We note that other groups (including Astakhov \etall, 2003 and
Astakhov \& Farrelly, 2004) have initiated bodies from the Hill sphere
without considering how this affects the resulting statistics, most
notably that many of their retrograde encounters are started on
already-stable orbits.  For future studies of capture near Jupiter, we
recommend the following procedure: 1) generate bodies starting from
1.0 $r_{H}$, 2) integrate each body backwards in time, 3) eliminate
those that remain near the planet when integrated backwards, and 4)
integrate the remaining bodies in normal time direction.  The benefits
of this approach over simply starting from 1.1 $r_H$ are that it
eliminates all stable-retrograde contamination while preserving high
number statistics.  Note, however, that this procedure still does not
allow a valid comparison between prograde and retrograde statistics.
To obtain the true ratio of captured progrades and retrogrades, it 
would probably be best to start the interlopers on heliocentric orbits.

\subsection{Scaling to unequal binary masses}

The impulse approximation (Eqs.~\ref{rtd} and~\ref{deltav}) implies
that a binary component's likelihood of capture depends only on 1) the
binary's tidal disruption radius and 2) the component's instantaneous
speed at the time of disruption.  Accordingly, for any mass ratio
$m_{1}:m_{2}$ and semi-major axis $a_B$, we can find an equal-mass
binary with the same $r_{td}$ and the same component speeds as $m_1$
and another equal-mass binary that matches these values for $m_2$.
Setting $r_{td}$ and $v$ equal for the two mass ratios, we solve
for the component mass, $m'$, and the semi-major axis, ${a_{B}}'$, of
the equal-mass binary matching $m_1$:

\begin{equation}
\label{mratio-m}
m'= \frac{4{m_{2}}^{3}}{(m_{1}+m_{2})^{2}}
\end{equation}

\noindent and

\begin{equation}
\label{mratio-a}
{a_{B}}' = a_{B}\left(\frac{2m_{2}}{m_1+m_2}\right) .
\end{equation}

\noindent The problem is symmetric, so for the equal-mass binary
matching $m_2$, we simply exchange $m_1$ and $m_2$ in the above
equations.

We tested these predictions for 3:2 and 4:1 mass ratios and display
the results in Table 1.  Each of the equal-mass components captures
with the same efficiency (to within 10$\%$) as either $m_1$ or $m_2$.
This is strong validation of the impulse approximation.  This sort of
scaling requires, of course, that the binary be split and therefore
applies only to our 1C and 2C-IND results.  By contrast, undisrupted
binaries (2C-BIN) behave as single objects and have a capture
efficiency that is independent of mass.

The impulse approximation allows us to make other predictions as
well.  For example, we can reverse the above scenario and predict the
capture rates of two equal-mass binaries corresponding to one
unequal-mass pair for which the capture rates of each component are
known.

Further, given the capture rate for a single equal-mass binary, we can
predict the capture efficiency of one component of an unequal-mass
pair, if one parameter of the unequal-mass binary ($m_1$, $m_2$, $m_1
+ m_2$, or $a_B$) is set.  This means that each equal-mass pair
matches to a whole family of unequal-mass binaries.

Finally, we can extend these techniques to guide large-scale
simulations.  Using the impulse approximation, and given equal-mass
capture statistics for all combinations of relevant masses and
separations, we can predict the capture rates for {\it all}
unequal-mass binaries -- with any mass ratio, total mass, and
separation.  Similarly, knowledge of unequal-mass capture rates for
one component of a binary with a fixed mass ratio allows us to
estimate the capture efficiencies of {\it all} equal-mass binaries.

Scaling with the impulse approximation is a powerful way to predict
capture rates.  Practically speaking, it means that capture of any
unequal-mass binaries can be predicted by studying just equal-mass
cases, and accordingly, we have restricted our numerical studies to
binaries with equal masses.

\subsection {Jupiter's eccentricity}

Most scenarios, including our own, propose that irregular satellite
capture occurred early in the Solar System's history
(Section~\ref{models}).  It is likely that Jupiter's eccentricity was
closer to zero at this time and its current value was obtained later.
Thus we have chosen to use $e_J$ = 0 in our simulations and believe it
is a reasonable assumption.

How would Jupiter's current eccentricity affect capture?  It is not
straightforward to extend this study to a non-zero jovian eccentricity
for the primary reason that the Jacobi constant is no longer a
constant of the motion.  Though Jupiter's current eccentricity is
small ($e_{J} \sim$ 0.048), it causes the calculated $C_J$ of an
interloper to vary by $\sim$0.2 over an orbit.  This is a variation
about ten times larger than our entire range of tested Jacobi
constants.  Thus we cannot simply assign an initial $C_J$ to binaries
or single objects approaching an eccentric Jupiter and compare the
capture results with the equivalent circular case.  Only for
eccentricities less than a hundredth of Jupiter's would the errors
introduced in the Jacobi constant be acceptable.

The results of Astakhov \& Farrelly (2004) suggest an order of
magnitude lower capture probability for single objects in the
eccentric case as compared to capture by a planet orbiting on a
circle.  However, their initial conditions were generated using the
same Jacobi constant values in the eccentric and the non-eccentric
cases.  Because of the variation in $C_J$ induced by eccentricity,
their method actually produces a much larger range of Jacobi constants
(and much higher approach speeds) for tests with an eccentric planet,
and the results should not be directly compared with the circular
case.  We believe their conclusions give artificially low
eccentric-case capture rates for this reason.

We do not expect Jupiter's eccentricity to strongly affect capture.
As the timescale for capture is much shorter than Jupiter's orbital
period, the planet's instantaneous location is the most relevant
factor.  At pericenter, Jupiter's Hill sphere is slightly smaller than
at its average distance from the Sun and encounter speeds are slightly
higher, likely resulting in fewer captures.  The opposite can be
expected at apocenter, causing the effects of eccentricity to average
out and probably produce little change in overall capture statistics.
To truly know the effects of the planet's eccentricity would require
large-scale integrations with the interlopers originating on
heliocentric orbits.  This is beyond the scope of our current work,
but again, we suspect the difference in capture rates will be small.

\subsection{Survivability of captured objects}

\label{survivability}

\begin{figure}[htp]
\plotone{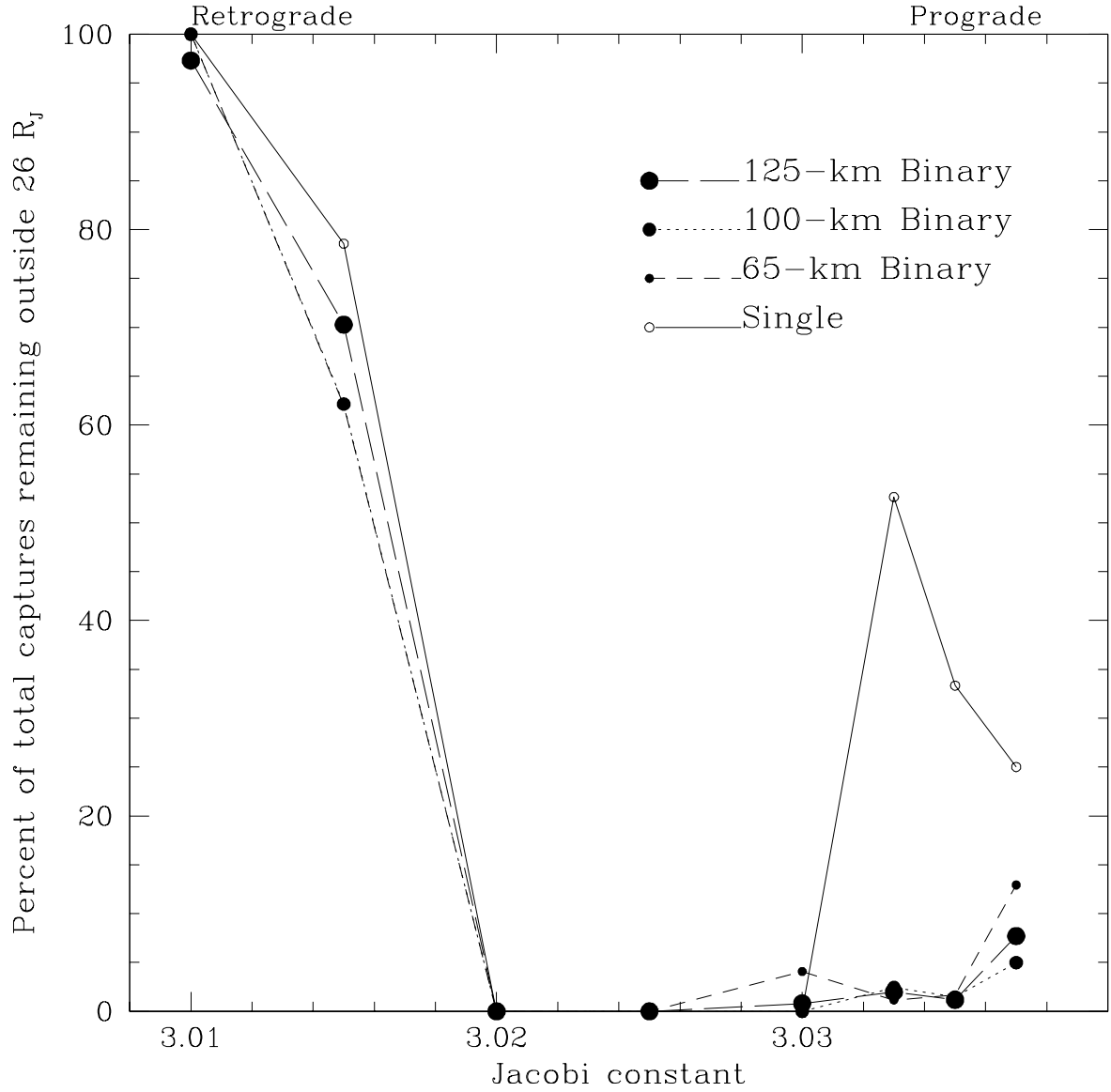}
\caption{The percent of captured objects that do not cross interior to
  26 $R_J$ (Callisto's semi-major axis) during the 1,000-year
  integrations.  The bodies were started on the Hill sphere.  Compare
  with the entire set of captures seen in Fig.~\ref{1rH}.}
\label{noCA}
\end{figure}

The post-capture orbits are initially very irregular and prone to
collisions with Jupiter's Galilean satellites (particularly its
outermost, Callisto) or the planet itself.  Figure~\ref{noCA} displays
the percent of captured bodies that are delivered onto orbits that do
not cross Callisto, which orbits Jupiter at 26 $R_J$.

An interesting feature is the lack of any captures without close
approaches for $C_{J}$ = 3.02 and 3.025.  These Jacobi constants
correspond to orbital inclinations near 90$\deg$ (see
Section~\ref{incl}), where the Kozai effect is strongest.  The Kozai
mechanism causes the orbits to become highly eccentric and subject to
collision with Jupiter or one of the Galilean satellites.  This is the
primary reason that no existing satellites in the Solar System have
inclinations near 90$\deg$.

All of the surviving binaries plotted here capture as 2C-BIN (though
not all 2C-BIN captures are survivors), largely because of the
separations we have studied.  The separations of these binaries are
optimized to give maximum capture percentages, leading to tidal
disruption radii that are very close to Jupiter (Eq.~\ref{rtd}).  Thus
these surviving binaries, which by definition have closest approaches
outside Callisto, do not cross $r_{td}$ and remain intact.  The
separations we have used, which are optimal for capture, do not
correspond with the maximum survival rate.  Somewhat larger
separations would result in a higher percent of surviving captures,
with a lower percent of overall captures.

The consequences of this are apparent in Figure~\ref{noCA}.  For $C_J
<$ 3.02, the curves for single objects and all three binary masses are
equivalent to within the error.  This is because most of the binary
captures in this $C_J$ range are 2C-BIN.  Since 2C-BIN captures act as
a single entity with no alteration from tidal disruption, the captured
population for $C_J <$ 3.02 is similar to that of the single bodies
and the fraction that avoid Callisto is also similar.  While the
surviving binaries for $C_J >$ 3.025 are still all 2C-BIN, 1C captures
dominate the 2C-BIN captures in this Jacobi constant range, but none
of the 1C captures are safe from Callisto.  Therefore, the percent of
the total binary captures that survive at high Jacobi constants is
very small.

Thus we find that while retrograde captures (low $C_J$) are rare (see
Fig.~\ref{1rH}), most of them are safe from collision with Callisto.
On the contrary, prograde captures from binaries (high $C_J$) are
numerous but prone to collision.  This is bad news for binary capture
-- the largest enhancements occur for captures that are most likely to
be removed by interactions with Callisto.  As we have already
discussed, examining larger separations will lead to a larger
percentage of capture orbits that are safe from collision.  This may
be a significant effect.  How else can collisions be avoided?

One possible way out is if the captures occurred before Callisto was
formed.  At these early times, a dense accretion disk surrounded
Jupiter, and the strong gas drag could have augmented capture rates
(as in Pollack \etall, 1979).  However, this is not compelling,
because satellites captured at this time would be prone to loss by
orbital decay and later by collisions with forming proto-satellites.

The most likely mechanism for preventing captured bodies from
colliding is gas drag from the remaining gas present outside
Callisto's orbit at the time of capture.  A small amount of gas is
necessary for our mechanism in order to shrink capture orbits to their
current sizes.  This process can also increase the pericenters of the
captured objects, causing them to avoid collision with Callisto.  A
typical collision timescale for a Callisto-crossing orbit with $a =
0.5r_{H}$ and i = 10$\deg$ is on the order of $10^6$ years, long
enough for gas to evolve the satellites onto safe orbits (see
Section~\ref{ourmodel}).  The timescale is longer for more tilted
orbits, but significantly shorter for retrograde orbits.

\begin{figure}[htp]
\plotone{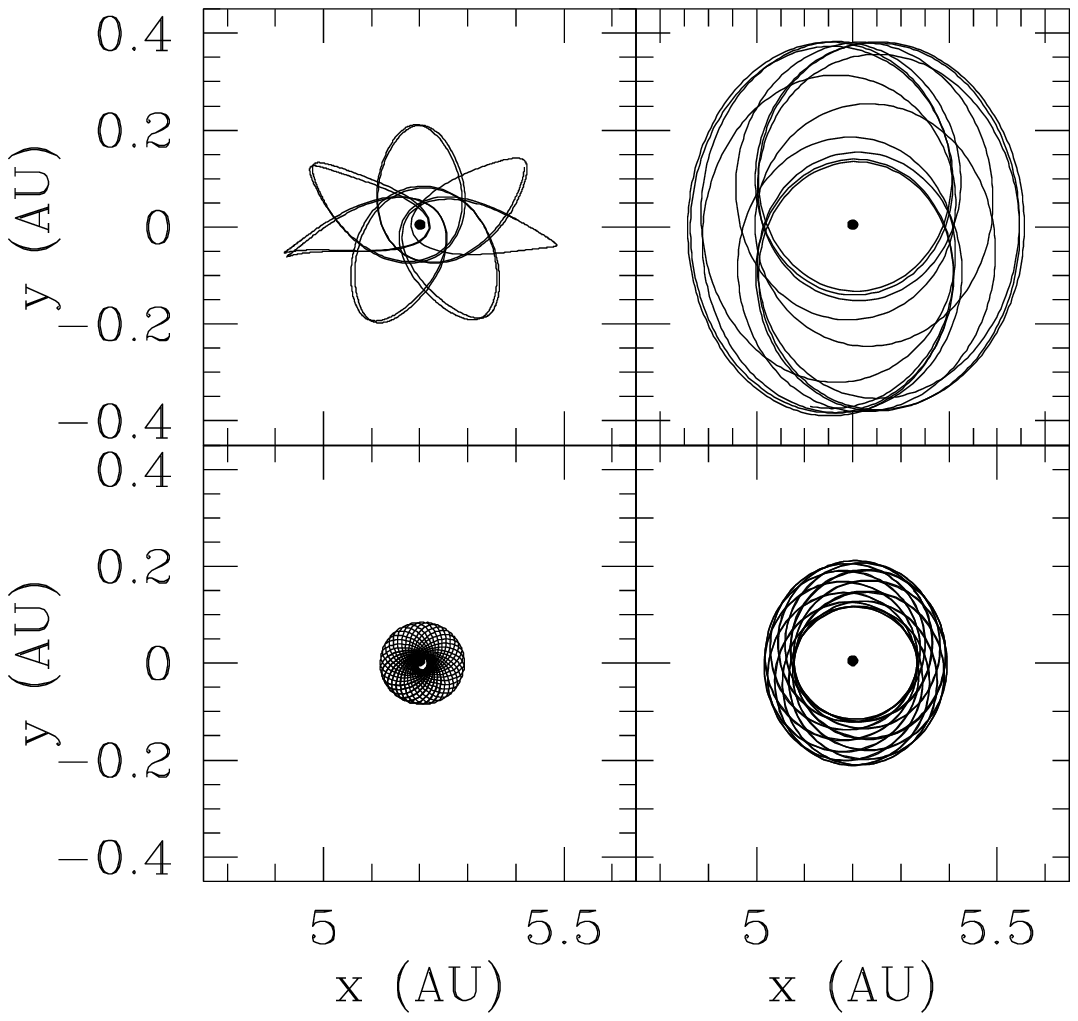}
\caption{An example of a simple gas drag applied to a prograde (left
  panels; the same initial orbit as in Figs.~\ref{cj3panel}
  and~\ref{exampleorbit}) and a retrograde orbit (right panels).  The
  orbits are shown immediately after capture in the upper panels, and
  the bottom panels show the orbits after 25,000 years of evolution.
  Jupiter is the dot in the center, and the Sun is to the left at
  (0,0).  These orbits are the result of 1C capture from equal-mass
  binaries with 65-km components.  Before disruption, the prograde
  binary had a separation of 70 $R_B$ and a Jacobi constant of $\sim$
  3.037.  The retrograde binary had an initial $C_J$ of $\sim$ 3.003
  and was initially separated by 460 $R_B$.}
\label{drag}
\end{figure}

A tenuous gas around Jupiter and Saturn near the end of planet
formation is consistent with our current understanding of planet
formation (Section~\ref{ourmodel}).  If gas was present at the time of
capture, its structure and density are not well constrained, and thus
we do not focus on the orbital evolution process itself in this work.
As a simple example, however, we simulated a drag force that acts
against the velocity vector and show that it is able to both shrink
the post-capture orbits and prevent the new satellites from collision.
In Fig.~\ref{drag}, we plot the initial and final states for a
prograde and retrograde orbit.  The gas drag is applied after capture
for simplicity; this is valid because the timescale for temporary
capture is $\sim$months, and the effects of the gas are negligible
over such a short time.  The final states are chosen so that the
orbits lie approximately where the current progrades and retrogrades
orbit at Jupiter (at $\sim\frac{1}{4} r_H$ and $\sim\frac{1}{2} r_H$,
respectively), and they have pericenters outside Callisto's orbit.
The drag strengths were set so that the evolution for both orbits
occurs over 25,000 years.  However, with a more tenuous gas than in
this example, the same evolution could take place on timescales 10-100
times longer.  The binary capture mechanism discussed here constrains
only the amount of orbital evolution, not the evolution rate, and
hence avoids the satellite loss problem of capture-by-gas-drag models.

\section{Discussion}

The new model discussed here, capture from low-mass binaries with
subsequent orbital evolution, has both significant advantages and
disadvantages in comparison to other suggested models.

One important advantage is that capture is viable at Jupiter and
Saturn, unlike three of the models discussed in Section~\ref{intro}.
For example, Agnor \& Hamilton's (2006) direct three-body capture
model works only for very large bodies in the gas giants'
high-approach-speed environments.  Also, the theory of Nesvorn{\'y}
\etal (2003) requires close approaches among the giant planets, of
which Jupiter has very few in the Nice model scenario.  Saturn
encounters its outer neighbors more frequently than Jupiter, but it
still suffers from few close approaches overall in this model.  In
Vokrouhlick{\'y} \etal (2007), the capture statistics from
planet-binary encounters are low for all planets, but especially
Jupiter and Saturn.  This is primarily because of the high relative
velocities assumed in their model.  Though these latter two capture
mechanisms in their current forms cannot explain the gas giants'
irregular satellites, they are worth further study, perhaps in the
context of altered versions of the Nice model or other early Solar
System models.

Our model also has an important advantage over that of Pollack \etall,
1979, in that our capture scenario allows for a much weaker gas, since
the gas here is needed for shrinking the orbits, not capturing satellites.
Also, the gas in our model can persist for much longer than that in
Pollack \etall, as a weaker gas does not have the problem of quickly
destroying the captured bodies.  Furthermore, various groups (e.g.,
Canup \& Ward, 2002) have proposed that the Galilean satellites formed
from a tenuous gas; if true, the gas was most likely even thinner when
the irregular satellites were captured.  Thus it is important that our
capture mechanism does not rely on a dense gas.

To directly compare our mechanism with that of Astakhov \etal (2003),
we need to consider both relative capture rates (and their
survivability) and the prevalence of binaries vs. single bodies.  In
our simulations, we find that binary capture can provide a significant
advantage over capturing from populations of single bodies for
binaries with particular characteristics: high enough masses
($\gtrsim$ 100 km), optimal separations ($\sim$10-20 $R_B$), and low
incoming energies (corresponding to Jacobi constants $\gtrsim$ 3.02
and mostly prograde encounters).  However, like Astakhov \etal (2003),
we also find that the probability of captures (from either binaries or
single bodies) of avoiding collisions with Callisto is low for readily
captured progenitors.  In Section~\ref{survivability}, we discussed
that this problem can be alleviated by altering the capture orbits
with the surrounding gas or by capturing binaries with
larger-than-optimal separations that do not lead to Callisto-crossing
orbits.

So, how common were easily-captured binaries early in Solar System
history?  This question is difficult to assess.  Observational surveys
of the current population of the cold, classical Kuiper belt
(i.e. objects with modest inclinations and eccentricities) find a 30\%
binary fraction among bodies larger than 100 km (Noll \etall, 2008),
many with nearly equal mass components.  Also, recent studies of
planetesimal formation have suggested that large, $\gtrsim$ 100-km
bodies may form quickly in the gaseous proto-planetary disk, providing
the building blocks of subsequent planet formation (Johansen \etall,
2007; Cuzzi \etall, 2008).  Binary formation is likely to be
contemporaneous with the formation of these bodies (Nesvorn{\'y},
2008).  Further, Morbidelli \etal (2009) have shown that the
size-frequency distribution of asteroids in the main belt is
consistent with large initial planetesimals, at least $\sim$100-km in
size.  Together these results indicate that the $\sim$100-km binary
objects considered here may have been quite common as the very last
portions of the Solar System's gas disk were being depleted.

Finally, the known irregular satellite population numbers $<$ 100, and
many of these are probably members of families -- thus, we need only
produce at most a few dozen captures.  While accounting for the origin
of such a small population is difficult, our simulations show that it
is likely binaries played a role.  We conclude by offering our model
as a new idea that alleviates many but not all of the problems faced
by previous models, but acknowledge that the without a detailed
understanding of the initial population including binary statistics, a
firm conclusion is not possible.

\clearpage

\begin{deluxetable}{cccc|ccc}
  \tablecolumns{7} \tabletypesize{\footnotesize} \tablecaption{Mass Ratio Tests For Binaries With $C_J$=3.03} 
  \tablewidth{0pt}
  \tablehead{ \colhead{Mass} &\colhead{Component} &\colhead{Total Binary} &\colhead{Separation} &\colhead{$m_1$ Capture} &\colhead{$m_2$ Capture} &\colhead{Total 1C Capture}\\ 
  \colhead{Ratio} &\colhead{Radii (km)} &\colhead{Mass ($10^{19}$kg)} &\colhead{(km)} &\colhead{Percentage} &\colhead{Percentage} &\colhead{Percentage} }
  \startdata
  3:2 & 113-98 & 2.0 & 1350 & {\bf 0.54} & {\it 0.96} & 1.50 \\
  1:1 & 85-85 & 1.0 & 1080 & {\bf 0.58} & {\bf 0.58} & 1.16 \\ 
  1:1 & 127-127 & 3.4 & 1620 & {\it 0.94} & {\it 0.94} & 1.88 \\
  \tableline
  4:1 & 124-78 & 2.0 & 1160 & {\bf 0.21} & {\it 1.17} & 1.38 \\
  1:1 & 42-42 & 0.13 & 465 & {\bf 0.24} & {\bf 0.24} & 0.48 \\
  1:1 & 169-169 & 8.1 & 1860 & {\it 1.29} & {\it 1.29} & 2.58 \\
  \tableline 
  \enddata 

  \renewcommand{\baselinestretch}{1} 

  \tablecomments{Two experiments with binaries of unequal mass.  The
  first four columns identify properties of the binary, while the
  final three columns list capture statistics (with $m_1 \geq m_2$).  We
  used the impulse approximation (Eqs.~\ref{rtd} and~\ref{deltav}) to
  determine the properties of equivalent equal-mass binaries that
  match the tidal disruption radius and speed of either $m_1$
  (boldfaced) or $m_2$ (italicized).  In all cases, our predicted
  percentages agree with the actual measurements to within about
  10$\%$.}
\label{massratios}
\end{deluxetable}

\clearpage

\renewcommand{\baselinestretch}{1}

\newpage


\begin{thebibliography}{}

\bibitem[none]{none} Agnor, C.~B. and Hamilton, D.~P., 2006a.
  Neptune's capture of its moon Triton in a binary-planet
  gravitational encounter.  Nature 441, 192-194.

\bibitem[none]{none} Agnor, C.~B. and Hamilton, D.~P., 2006b.
  Satellite capture via binary-planet gravitational encounters.
  Bull. Am. Astron. Soc. 38, 674.

\bibitem[none]{none} Astakhov, S.~A. and Farrelly, D., 2004.  Capture
  and escape in the elliptic restricted three-body problem.
  Mon. Not. R. Astron. Soc.  354, 971-979.

\bibitem[none]{none} Astakhov, S.~A., Burbanks, A.~D., Wiggins, S.,
  and Farrelly, D., 2003.  Chaos-assisted capture of irregular moons.
  Nature 423, 264-267

\bibitem[none]{none} Bate, M.~R., Lubow, S.~H., Ogilvie, G.~I., and
  Miller, K.~A., 2003.  Three-dimensional calculations of high- and
  low-mass planets embedded in protoplanetary discs.
  Mon. Not. R. Astron. Soc. 341, 213-229.

\bibitem[none]{none} Canup, R.~M. and Ward, W.~R., 2002.  Formation of
  the Galilean satellites: Conditions of accretion.  Astron. J. 124,
  3404-3423.

\bibitem[none]{none} Cieza, L. and 19 colleagues, 2007.  The Spitzer
  c2d survey of weak-line T Tauri stars. II. New constraints on the
  timescale for planet building.  Astrophys. J. 667, 308-328.

\bibitem[none]{none} Colombo, G. and Franklin, F.~A., 1971.  On the
  formation of the outer satellite groups of Jupiter.  Icarus 15,
  186-189.

\bibitem[none]{none} {\'C}uk, M. and Burns, J.~A., 2004a.  Gas-drag-assisted
 capture of Himalia's family.  Icarus 167, 369-381.

\bibitem[none]{none} {\'C}uk, M. and Burns, J.~A., 2004b.  On the secular
  behavior of irregular satellites.  Astron. J. 128, 2518-2541.

\bibitem[none]{none} Cuzzi, J.~N., Hogan, R.~C., and Shariff, K.,
  2008.  Toward planetesimals: Dense chondrule clumps in the
  protoplanetary nebula.  Astrophys. J. 687, 1432-1447.

\bibitem[none]{none} D'Angelo, G., Henning, T., and Kley, W., 2003.
  Thermohydrodynamics of circumstellar disks with high-mass planets.
  Astrophys. J. 599, 548-576.

\bibitem[none]{none} Hamilton, D.~P. and Burns, J.~A., 1991.  Orbital
  stability zones about asteroids.  Icarus 92, 118-131.

\bibitem[none]{none} Henon, M., 1969.  Numerical exploration of the
  resticted problem, V.  Astron. Astrophys. 1, 223-238.

\bibitem[none]{none} Heppenheimer, T.~A. and Porco, C., 1977.  New
  contributions to the problem of capture.  Icarus 30,
  385-401.

\bibitem[none]{none} Jewitt, D. and Sheppard, S., 2005.  Irregular
  satellites in the context of planet formation.  Space Science
  Reviews 116, 441-455.

\bibitem[none]{none} Johansen, A., Oishi, J.~S., Low, M.-M.~M., Klahr,
  H., Henning, T., and Youdin, A., 2007.  Rapid planetesimal formation
  in turbulent circumstellar disks.  Nature 448, 1022-1025.

\bibitem[none]{none} Lubow, S.~H., Seibert, M., and Artymowicz, P., 1999.
Disk accretion onto high-mass planets.  Astrophys. J. 526, 1001-1012.

\bibitem[none]{none} Merline, W.~J. and 11 colleagues, 2007.  The
  search for Trojan binaries.  Bull. Am. Astron. Soc. 38, 538.

\bibitem[none]{none} Morbidelli, A., Bottke, W.~F., Nesvorn{\'y}, D., and
  Levison, H.~F., 2009.  Asteroids were born big.  Icarus, in press.

\bibitem[none]{none} Mosqueira, I. and Estrada, P.~R., 2003.
  Formation of the regular satellites of giant planets in an extended
  gaseous nebula I: Subnebula model and accretion of satellites.  Icarus
  163, 198-231.

\bibitem[none]{none} Murray, C.~D. and Dermott, S.~F., 1999.  Solar
  System Dynamics.  Cambridge Univ. Press, Cambridge, UK.

\bibitem[none]{none} Nesvorn{\'y}, D., 2008.  Formation of Kuiper belt
  binaries.  Bull. Am. Astron. Soc. 40, 464.

\bibitem[none]{none} Nesvorn{\'y}, D., Alvarellos, J.~L.~A., Dones, L.,
  and Levison, H.~F., 2003.  Orbital and collisional evolution of the
  irregular satellites.  Astron. J. 126, 398-429.

\bibitem[none]{none} Nesvorn{\'y}, D., Vokrouhlick{\'y}, D., and Morbidelli,
  A., 2007.  Capture of irregular satellites during planetary
  encounters.  Astron. J. 133, 1962-1976.

\bibitem[none]{none} Noll, K.~S., Grundy, W.~M., Stephens, D.~C.,
  Levison, H.~F., Kern, S.~D., 2008.  Evidence for two populations of
  classical transneptunian objects: The strong inclination dependence of
  classical binaries.  Icarus 194, 758-768.

\bibitem[none]{none} Papaloizou, J.~C.~B., Nelson, R.~P., Kley, W.,
  Masset, F.~S., and Artymowicz, P., 2007.  Disk-planet interactions
  during planet formation.  In: Reipurth, D., Jewitt, D., and Keil,
  K. (Eds.), Protostars and Planets V, Univ. of Arizona Press, Tucson,
  655-668.

\bibitem[none]{none} Pollack, J.~B., Burns, J.~A., and Tauber, M.~E.,
  1979.  Gas drag in primordial circumplanetary envelopes -- a
  mechanism for satellite capture.  Icarus 37, 587-611.

\bibitem[none]{none} Pollack, J.~B., Hubickyj, O., Bodenheimer, P.,
  Lissauer, J.~J., Podolak, M., and Greenzweig, Y., 1996.  Formation of
  the giant planets by concurrent accretion of solids and gas.  Icarus
  124, 62-85.

\bibitem[none]{none} Rauch, K.~P. and Hamilton, D.~P., 2002.  The
  HNBody package for symplectic integration of nearly-Keplerian
  systems.  Bull. Am. Astron. Soc. 34, 938.

\bibitem[none]{none} Saha, P. and Tremaine, S., 1993.  The orbits of the
  retrograde Jovian satellites.  Icarus 106, 549-562.
 
\bibitem[none]{none} Sheppard, S.~S. and Jewitt, D.~C., 2003.  An
  abundant population of small irregular satellites around Jupiter.
  Nature 423, 261-263.

\bibitem[none]{none} Silverstone, M.~D. and 16 colleagues, 2006.
  Formation and Evolution of Planetary Systems (FEPS): Primordial warm
  dust evolution from 3 to 30 Myr around Sun-like stars.
  Astrophys. J. 639, 1138-1146.

\bibitem[none]{none} Strutskie, M.~F., Dutkevitch, D., Strom, S.~E.,
  Edwards, S., Strom, K.~M., and Shure, M.~A., 1990.  A sensitive
  10-micron search for emission arising from circumstellar dust
  associated with solar-type pre-main-sequence stars.  Astron. J. 99,
  1187-1195.

\bibitem[none]{none} Tsiganis, K., Gomes, R., Morbidelli, A., and
  Levison, H.~F., 2005.  Origin of the orbital architecture of the
  giant planets of the Solar System.  Nature 435, 459-461.

\bibitem[none]{none} Whipple, A.~L. and Shelus, P.~J., 1993.  A
  secular resonance between Jupiter and its eighth satellite?  Icarus
  101, 265-271.

\bibitem[none]{none} Vokrouhlick{\'y}, D., Nesvorn{\'y}, D., and Levison,
  H.~F., 2008.  Irregular satellite capture by exchange reactions.
  Astron. J. 136, 1463-1476.

\end{thebibliography}
\end{document}